\documentclass{jfm}

\usepackage{graphicx}
\usepackage{epstopdf,epsfig}
\usepackage{newtxtext}
\usepackage{newtxmath}
\usepackage{natbib}
\usepackage{hyperref}
\usepackage{longtable}
\usepackage{ragged2e} 
\usepackage{booktabs,makecell, multirow, tabularx}

\hypersetup{
    colorlinks = true,
    urlcolor   = blue,
    citecolor  = black,
}

\newcommand{\RomanNumeralCaps}[1]

\usepackage{soul}
\usepackage{xcolor}
\soulregister{\ref}{1} 
\soulregister{\citep}{7} 
\usepackage{tikz}
\usepackage{caption}

\title{Particle transport and deposition in wall-sheared thermal turbulence}

\author{Ao Xu\aff{1}$^,$\aff{2}$^,$\aff{3},
        Ben-Rui Xu\aff{1}
        \and Heng-Dong Xi\aff{1}$^,$\aff{2}$^,$\aff{3} \corresp{\email{hengdongxi@nwpu.edu.cn}}}

\affiliation{
\aff{1}School of Aeronautics, Northwestern Polytechnical University, Xi'an 710072, PR China
\aff{2}Institute of Extreme Mechanics, Northwestern Polytechnical University, Xi'an 710072, PR China
\aff{3}National Key Laboratory of Aircraft Configuration Design, Xi'an 710072, PR China
}

\begin{document}
\maketitle

\begin{abstract}
We studied the transport and deposition behaviour of point particles in Rayleigh-B\'enard convection cells subjected to Couette-type wall shear. 
Direct numerical simulations (DNSs) are performed for Rayleigh number ($Ra$) in the range $10^{7} \leq Ra \leq 10^9$ with a fixed Prandtl number $Pr = 0.71$, while the  wall-shear Reynolds number ($Re_w$) is in the range $ 0 \leq Re_w \leq 12000$. 
With the increase of $Re_w$, the large-scale rolls expanded horizontally, evolving into zonal flow  in two-dimensional simulations or streamwise-oriented rolls in three-dimensional simulations.
We observed that, for particles with a small Stokes number ($St$), they either circulated within the large-scale rolls when buoyancy dominated or drifted near the walls when shear dominated. 
For medium $St$ particles, pronounced spatial inhomogeneity and preferential concentration were observed regardless of the prevailing flow state. 
For large $St$ particles, the turbulent flow structure had a minor influence on the particles’ motion; although clustering still occurred, wall shear had a negligible influence compared with that for medium $St$ particles. 
We then presented the settling curves to quantify the particle deposition ratio on the walls. 
Our DNS results aligned well with previous theoretical predictions, which state that small $St$ particles settle with an exponential deposition ratio and large $St$ particles settle with a linear deposition ratio. 
For medium $St$ particles, where complex particle-turbulence interaction emerges, we developed a new model describing the settling process with an initial linear stage followed by a nonlinear stage. 
Unknown parameters in our model can be determined either by fitting the settling curves or using empirical relations. 
Compared with DNS results, our model also accurately predicts the average residence time across a wide range of $St$ for various $Re_w$.
\footnote{
This article may be downloaded for personal use only.
Any other use requires prior permission of the author and Cambridge University Press.
This article appeared in Xu \emph{et al.}, J. Fluid Mech. \textbf{999}, A15 (2024) and may be found at \url{https://doi.org/10.1017/jfm.2024.936}.
}
\end{abstract}

\begin{keywords}
B\'enard convection, particle/fluid flow, turbulent convection
\end{keywords}


\section{Introduction}
\label{sec:1 Introduction}
Turbulence significantly affects the dynamics of particulate suspensions, as evidenced ubiquitously in nature and daily life \citep{guha2008transport,toschi2009lagrangian,mathai2020bubbly,brandt2022particle}. 
For example, aeolian processes can lead to the airborne movement or saltation of sand grains of different sizes under the influence of wind, triggering dust storms, especially under unstable thermal conditions \citep{zheng2009mechanics,zhang2024structure}. 
Another example is the use of heating, ventilation and air conditioning systems, which can extend the suspension duration and spread of respiratory droplets, with this effect amplified by air currents in indoor environments \citep{mittal2020flow,hedworth2021mitigation}. 
In these scenarios, buoyancy forces generated by temperature differences drive turbulent flows. 
To study thermal turbulence, the Rayleigh-B\'enard (RB) convection system serves as a canonical paradigm. 
This system involves a fluid layer heated from the bottom and cooled from the top \citep{ahlers2009heat,lohse2010small,chilla2012new,wang2020vibration,xia2023tuning}. 
The control parameters in the RB system include the Prandtl number ($Pr$), which describes a fluid's thermophysical properties, and the Rayleigh number ($Ra$), which describes the relative strength of buoyancy forces vs thermal and viscous dissipation. 
In RB turbulent convection, ubiquitous coherent structures include small-scale plumes and large-scale circulation (LSC). 
After detaching from the boundary layers, sheet-like plumes transform into mushroom-like plumes through mixing and clustering \citep{zhou2007morphological}. 
Through plume-vortex and plume-plume interactions, thermal plumes further self-organize into the LSC, which spans the whole convection cell \citep{xi2004laminar}.

Due to these complex multi-scale coherent structures of thermal turbulence, efforts have been devoted to investigating the transport and deposition behaviour of particles in thermal convection. 
A particularly interesting aspect is the spatial distribution and deposition rate of particles in thermal turbulence. 
Depending on the Stokes number ($St$), which describes the particle inertia relative to that of the fluid, the particles' dynamic behaviour can be classified into three categories.
For particles with a small $St$, they are randomly distributed and behave like tracer particles; in the limit of infinitely small $St$, they exhibit an exponential deposition rate \citep{martin1989fluid}. 
For particles with a large $St$, their motion is almost unaffected by the underlying thermal turbulence; in the limit of infinitely large $St$, they settle at a constant speed of terminal velocity $v_{t}$, and the deposition rate on the wall follows a linear law as derived from Stokes' law. 
For particles with a medium $St$, they tend to cluster into band-like structures, and these structures are found to align with the vertical movement of plumes in thermal turbulence \citep{park2018rayleigh}. 
To predict the deposition rate for medium $St$, \citet{patovcka2020settling,patovcka2022residence} developed a mathematical model that describes particle sedimentation as a stochastic process. 
Particles move from areas of intense convection to lower-velocity regions near the horizontal boundaries of the cell, with the possibility of escaping low-velocity regions without settling. 
In addition, the effects of thermal and mechanical coupling on the turbulence structure of the LSC and boundary layers, as well as particles’ motion, have received wide interest recently through simulations of two-way coupling \citep{oresta2013effects,park2018rayleigh} or four-way coupling \citep{demou2022turbulent} between the fluid and particles.
Recent progress includes the works of \citet{du2022wall}, \citet{yang2022energy,yang2022dynamic}, \citet{sun2024modulation} and \citet{chen2024particle}, which incorporate thermal conduction and radiation, accounting for the thermal backreaction on the flow and examining its implications for heat transfer efficiency and flow structure modulation.

In addressing the complexities of natural and engineering fluid systems, such as atmospheric convection, oceanic currents and indoor air ventilation, it is crucial to examine the interactions between vertical buoyancy and the horizontal shear force \citep{hori2023jupiter}. 
For example, \citet{blass2020flow,blass2021effect} incorporated a Couette-type shear into the RB system, where the top and bottom walls move in opposite directions at a constant speed. 
They found a transition in the flow dynamics from buoyancy-dominated to shear-dominated regimes as the strength of wall shear increased. 
In the buoyancy-dominated regime, the flow structure resembles canonical RB convection. 
In the shear-dominated regime, they observed the development of large meandering rolls. 
\citet{yerragolam2022small} then analysed the spectra of convective flux and turbulent kinetic energy, thereby offering insights into the small-scale flow structures within the same convection cell. 
Further, \citet{jin2022shear} reported enhanced interactions between the LSC and secondary flows over rough shearing surfaces, leading to an increased generation of thermal plumes. 
\citet{xu2023wall} manipulated the movement of adiabatic sidewalls, thereby inducing vertical fluid motion that enhances heat transfer efficiency and can cause turbulent flows to relaminarize. 
In double diffusive convection, \citet{li2022flow} observed that even weak shear significantly alters the fingering morphology and transport properties of the system. 
Although recent advancements have shed light on the dynamics of single-phase and wall-sheared thermal turbulence, the complex interactions between a dispersed particulate phase and its surrounding fluid in wall-sheared thermal convection remain largely unexplored.

In this work, we aim to investigate particle transport and deposition behaviour within thermal turbulence under the influence of wall shearing. 
Motivated by studies of sand grains and pollutants dispersed in the air, we have chosen particles that are micrometres in size and much heavier than the surrounding fluid. 
Existing studies on particle-laden RB turbulence have largely focused on particles circulating within the LSC; here, we additionally consider scenarios where the flow structure changes under the influence of horizontal shear forces. 
The rest of this paper is organized as follows. 
In $\S$ \ref{sec:2 Numerical}, we present numerical details for the direct numerical simulation (DNS) of wall-sheared thermal turbulence laden with point particles. 
In $\S$ \ref{sec:3 Results}, we analyse the particles' transport behaviour in the cell and their spatial distribution, followed by their deposition patterns and a mathematical model to describe particle deposition rates. 
In $\S$ \ref{sec:4 Conclusion}, the main findings of the present work are summarized.

\section{Numerical methods}
\label{sec:2 Numerical}

\subsection {Problem statement }
\begin{figure}
	\centerline{\includegraphics[width=0.99\textwidth]{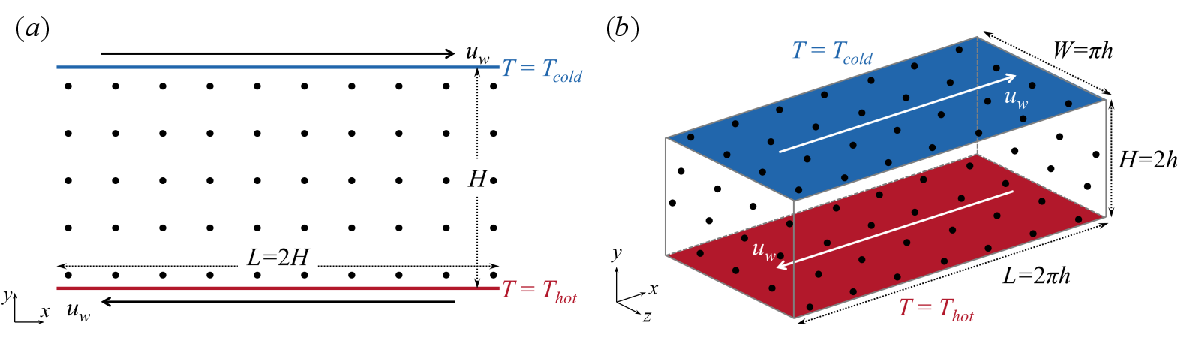}}
	\caption{Schematic illustration of the wall-sheared convection cell and particles' initial positions for:  (\emph{a}) the 2-D simulations; (\emph{b}) the 3-D simulations.}
	\label{fig:demo}
\end{figure}

We explore the dynamics of particles within a two-dimensional (2-D) convection cell of dimensions $L \times H$ (see figure \ref{fig:demo}\emph{a})  and within a three-dimensional (3-D) convection cell of dimensions $L \times H \times W$ (see figure \ref{fig:demo}\emph{b}). 
Here, $L$ is the length, $H$ is the height and $W$ is the width of the simulation domain, $x$ represents the wall-shear direction, $y$ represents the wall-normal direction and $z$ represents the spanwise direction.
The top and bottom walls of the cell are maintained at constant temperatures of $T_{\emph{\text{cold}}}$ and $T_{\emph{\text{hot}}}$, respectively, while they move in opposite directions at a constant speed of $u_w$. 
The cell’s vertical walls are configured to be periodic. 
Our simulation protocol is as follows:
we start the simulation of single-phase turbulent thermal convection, and particles are introduced into the turbulence once the flow reaches a statistically stationary state. 
Initially, the cell contains $N_{0}$ uniformly distributed stationary particles.  
We then advance the fluid flow and particle motion simultaneously, collecting statistics.
We assume that particles adhere to the wall upon contact, ceasing further movement within the convection cell and no longer interacting with other deposited particles. 

\subsection {Direct numerical simulation of thermal turbulence}
In incompressible thermal convection, we employ the Boussinesq approximation to account for the temperature as an active scalar influencing the velocity field via buoyancy effects, under the assumption of constant transport coefficients. 
The equations governing fluid flow and the heat transfer process can be written as
\begin{equation}\label{massequation}
\nabla \cdot \boldsymbol{u}_f = 0
\end{equation}
\begin{equation}\label{momentumequation}
\frac{\partial \boldsymbol{u}_f}{\partial t} + \boldsymbol{u}_f \cdot \nabla \boldsymbol{u}_f = 
- \frac{1}{\rho_0} \nabla P + \nu \nabla^2 \boldsymbol{u}_f + g\beta (T_f - T_0) \hat{\boldsymbol{y}}
\end{equation}
\begin{equation}\label{energyequation}
\frac{\partial T_f}{\partial t} + \boldsymbol{u}_f \cdot \nabla T_f = \alpha \nabla^2 T_f
\end{equation}
Here, ${{\boldsymbol{u}}_f}$ is the fluid velocity, and  $P$ and ${T_f}$  are the pressure and temperature of the fluid, respectively. 
The coefficients $\beta $, $\nu $ and $\alpha $  denote the thermal expansion coefficient, kinematic viscosity and thermal diffusivity, respectively. 
Reference state variables are indicated by subscript zeros.  
Also, $g$ represents gravitational acceleration, and  $\hat{\boldsymbol{y}}$ denotes the unit vector parallel to gravity  in the wall-normal direction.

Introducing the non-dimensional variables
\begin{equation}
\begin{aligned}
    & \boldsymbol{x}^* = \boldsymbol{x}/H, \quad t^* = t/ \sqrt{H/(g\beta \Delta_T)}, \quad \boldsymbol{u}_f^* = \boldsymbol{u}_f / \sqrt{g\beta \Delta_T H}, \\
    & P^* = P/ (\rho_0 g\beta \Delta_T H), \quad T_f^* = (T_f - T_0)/\Delta_T
\end{aligned}
\end{equation}
we can rewrite (\ref{massequation})-(\ref{energyequation}) in a dimensionless form as
\begin{equation}
\nabla \cdot \boldsymbol{u}_f^* = 0
\end{equation}
\begin{equation}
\frac{\partial \boldsymbol{u}_f^*}{\partial t^*} + \boldsymbol{u}_f^* \cdot \nabla \boldsymbol{u}_f^* = -\nabla P^* + \sqrt{\frac{Pr}{Ra}} \nabla^2 \boldsymbol{u}_f^* + T_f^* \hat{\boldsymbol{y}}
\end{equation}
\begin{equation}
\frac{\partial T_f^*}{\partial t^*} + \boldsymbol{u}_f^* \cdot \nabla T_f^* = \sqrt{\frac{1}{PrRa}} \nabla^2 T_f^*
\end{equation}
Here, $H$  denotes the cell height and $\Delta_T$  the temperature difference between the heating and cooling walls. 
The Rayleigh number ($Ra$) and the Prandtl number ($Pr$) are defined as
\begin{equation}
Ra = \frac{g \beta \Delta_T H^3}{\nu_f \alpha}, \quad Pr = \frac{\nu_f}{\alpha}
\end{equation}
When an external wall shear is introduced, an additional control parameter, the wall-shear Reynolds number ($Re_{w} = H{u_w}/\nu $), is needed. 
The competition between buoyancy and shear effects can be quantified by  the bulk Richardson number as $Ri = Ra/(Re_w^2Pr)$ \citep{blass2020flow,blass2021effect,yerragolam2022small}.

We adopt the spectral element method implemented in the open-source Nek5000 solver (version v19.0) for our DNS. 
Further details on the spectral element method and the validation of the Nek5000 solver are available in \citet{fischer1997overlapping} and \cite{kooij2018comparison}. 
Previously, we verified the Nek5000 simulation results for wall-sheared thermal convection against those from an in-house  GPU solver based on the lattice Boltzmann method \citep{xu2017accelerated,xu2023multi}.
The results from both the Nek5000 solver and our lattice Boltzmann solver demonstrated consistent results for wall-sheared thermal convection \citep{xu2023wall}.

\subsection {Kinematics of the particles}
We consider particles that are small enough not to influence the turbulence structure. 
The forces acting on a particle include the gravitational force $\boldsymbol{F}_\text{gravity}$ and the drag force $\boldsymbol{F}_\text{drag}$. 
The gravitational force $\boldsymbol{F}_\text{gravity}$ points in the direction of gravitational acceleration, counteracted by buoyancy acting in the opposite direction, which is given as
\begin{equation}
\boldsymbol{F}_\text{gravity} = \rho_p V_p \boldsymbol{g} - \rho_f V_p \boldsymbol{g}
\end{equation}
where $\rho _p$  and $V_p$ denote the particles’ density and volume, respectively. 
The drag force $\boldsymbol{F}_\text{drag}$ results from the particles’ effort to match the velocity of surrounding fluid, and is given as
\begin{equation}
\boldsymbol{F}_\text{drag} = \frac{m_p}{\tau_p} \left( \boldsymbol{u}_f - \boldsymbol{u}_p \right) f(Re_p)
\end{equation}
Here,  $\boldsymbol{u}_f$ is the fluid velocity at the particle's location, which is obtained by interpolating the fluid velocity fields from the mesh to the particle locations.
The parameters $m_p$ and $\boldsymbol{u}_p$ represent the mass and velocity of the particle, respectively. 
The particle response time is $\tau_p = \rho _pd_p^2/(18\mu_f)$, where $d_p$ is the particle diameter. 
The drag correction factor $f(Re_p)$ is a function of the particle Reynolds number  $Re_p = d_p|\boldsymbol{u}_f - \boldsymbol{u}_p|/\nu_f$. 
We determined $f(R{e_p})$ in our simulations by dynamically calculating $Re_{p}$ for each particle and then applying the correction factor based on the formula $f(Re_{p}) = 1 + 0.15Re_{p}^{0.687}$ \citep{clift1978bubbles}.
In cases where $Re_{p} \ll 1$, we have $f(Re_{p}) \approx 1$, indicating the validity of Stokes' drag law.
In this work, the 2-D cases can be approximated as a cross-section of the convection cell.
Similar to \citet{yang2022energy,yang2022dynamic}, we assume the fluid layer has a finite but very thin thickness $d_p$ in the spanwise direction. 
Both the fluid and particle motions are restricted in the spanwise direction, preventing movement along this axis. 
This restriction reduces the degrees of freedom, simplifying the dynamics. 
Consequently, while the overall flow is predominantly two-dimensional, the particles experience a 3-D environment due to the finite thickness of the fluid layer.
The same formulation for the particle response time has been used by \citet{patovcka2020settling} and \citet{yang2022energy,yang2022dynamic}  in their 2-D studies.

The particle motion is then described by \citep{van2008numerical,maxey2017simulation}
\begin{equation}
\frac{d \boldsymbol{x}_p(t)}{dt} = \boldsymbol{u}_p(t)
\end{equation}
\begin{equation}
\frac{d \boldsymbol{u}_p(t)}{dt} = \frac{\rho_p - \rho_f}{\rho_p} \boldsymbol{g} + \frac{\boldsymbol{u}_f(t) - \boldsymbol{u}_p(t)}{\tau_p} f(Re_p)
\end{equation}
Here, ${{\boldsymbol{x}}_p}$ denotes the position of the particle. 
For simplicity, we only consider the translational motion of isotropic spherical particles. 
To capture the anisotropic behaviour of particles in turbulent flows, the orientation and rotation dynamics of inertial particles can be modelled as described by \citet{challabotla2015orientation} and \citet{calzavarini2020anisotropic}. 
After non-dimensionalizing with the scales introduced above, we obtain
\begin{equation}
\frac{d \boldsymbol{x}_p^*}{dt^*} = \boldsymbol{u}_p^*
\end{equation}
\begin{equation} \label{eq:particle}
\frac{d \boldsymbol{u}_p^*}{dt^*} = -\Lambda \hat{\boldsymbol{y}} + \frac{\boldsymbol{u}_f^* - \boldsymbol{u}_p^*}{St_f} f(Re_p)
\end{equation}
Here, $\Lambda  = \left( {{\rho _p} - {\rho _f}} \right)/\left( {{\rho _p}\beta {\Delta _T}} \right)$  denotes the buoyancy ratio, which describes the relative importance of particle buoyancy with respect to thermal buoyancy of the fluid and  
$S{t_f} = {\tau _p}/{t_f}$ denotes the particle Stokes number, which describes the particle response time ${\tau _p}$ relative to the flow time scale quantified by the free-fall time ${t_f} = \sqrt {H/(g\beta {\Delta _T})}$. 
 At small Stokes numbers, the particle response time $\tau_p$ is very small, making the term $f(Re_p)(\mathbf{u}_f - \mathbf{u}_p)/\tau_p$ large unless $\mathbf{u}_p$ is very close to $\mathbf{u}_f$. 
As a result, the particle velocity $\mathbf{u}_p$ quickly matches the fluid velocity $\mathbf{u}_f$. 
In this regime, the drag force term dominates the particles' motion, and gravitational effects are minimal as particles follow the fluid flow.
At large Stokes numbers, the particle response time $\tau_p$ is large, causing the term $f(Re_p)(\mathbf{u}_f - \mathbf{u}_p)/\tau_p$ to become small. 
Hence, the drag force term is negligible, and the particle motion is dominated by the gravitational term $(\rho_p - \rho_f)\mathbf{g}/\rho_p$. 
In this regime, particles sediment with a velocity that increases linearly over time as the effect of gravity becomes more significant.

To solve the above differential equations that describe the Lagrangian particle motion, we adopt a particle-in-cell (PIC) library implemented in ppiclF \citep{zwick2019ppiclf}.
The PIC library, ppiclF, is integrated into Nek5000 by establishing a communication and data exchange interface between the two codes. 
This process involves initializing the particle system within the Nek5000 framework and ensuring effective communication between the fluid solver and the particle solver.
Each particle’s position and velocity are integrated separately. 
To compute the fluid velocities at particle locations, the library uses spectral polynomial interpolation.
This involves mapping Gauss-Lobatto-Legendre points within each spectral elements and evaluating barycentric Lagrange polynomials at the particles' coordinates, enhancing both accuracy and computational efficiency.

For particles settling in a stationary fluid where $\mathbf{u}_{f}=0$, at terminal settling velocity $\mathbf{u}_{p}(t)=\mathbf{u}_{t}$, the particle's acceleration is zero (i.e. $d\mathbf{u}_{p}/dt=0$). 
The speed of terminal velocity $v_{t}$ is then given by
\begin{equation}
v_t = \frac{d_p^2 (\rho_p - \rho_f) g}{18 \mu_f} \frac{1}{f(Re_p)}
\end{equation}
Due to the nonlinear relationship $f\left(Re_{p}(v_{t})\right)=1+0.15(d_{p}v_{t}/\nu_{f})^{0.687}$, we cannot obtain an analytical solution directly.
Instead, we use iterative methods to solve for $v_{t}$ from the above equation.

\subsection {Simulation settings}
We present 2-D simulation results  in the cell with $L=2H$ for Rayleigh numbers in the range of  $10^{7} \leq Ra \leq 10^9$ and a fixed Prandtl number of $Pr = 0.71$ (motivated by the studies of sand grains and pollutants dispersed in the air), while the  wall-shear Reynolds number $Re_w$ is in the range $0 \leq Re_w \leq 12000$. 
To demonstrate the relevance of our results for 3-D wall-sheared thermal convection, we conducted a set of 3-D simulations in a cell with dimensions $L\times H \times W=2\pi h \times 2h \times \pi h$ at $Ra=10^{7}$ and $Pr=0.71$. 
Here, $h=H/2$ is the half-height of the cell.
The wall-shear Reynolds number for the 3-D simulations is in the range $0 \leq Re_{w} \leq 4000$.
Recently, \citet{jie2022existence}, \citet{motoori2022role}, \citet{gao2023direct} and \citet{zhang2023electrostatic} found that this domain size can accurately capture particle behaviours.
The properties of air at a reference temperature of 300 K, including density, thermal expansion coefficient and kinematic viscosity, are listed in table \ref{tab:parameter}.
Additionally, we provide the height of the convection cell for a temperature difference of 5 K.
\begin{table}
	\begin{center}
		\def~{\hphantom{0}}
		\begin{tabular}{cc}
			Parameter	&  Value\\
			Rayleigh number ($Ra$)	&  $10^{7}, 10^8, 10^{9}$\\
			Prandtl number ($Pr$)	&  0.71\\
			Reference temperature ($T_0$)	&  300 K\\
			Reference fluid density (${\rho _0}$)	&  1.18 kg/m$^3$\\
			Thermal expansion coefficient ($\beta $)	&  $3.36\times10^{-3}$ K$^{-1}$\\
			Kinematic viscosity (${\nu_f}$)	& $1.57\times10^{-5}$ m$^2$/s \\
			Thermal diffusivity (${\alpha_f}$)	& $2.22\times10^{-5}$ m$^2$/s \\
			Temperature differences (${\Delta_T}$)	&  5 K\\
			Cell height ($H$)	&  0.28 m, 0.60 m, 1.28 m \\
		\end{tabular}%
		\caption{Simulation parameters and the corresponding fluid properties.}
		\label{tab:parameter}
	\end{center}
\end{table}

In the Nek5000 solver, the effective grid number is determined by the product of the number of spectral elements and the polynomial order. 
 For the 2-D simulations, the cases with $Ra = 10^{7}$ and $10^{8}$ use uniform grids, while the cases with $Ra = 10^{9}$ use non-uniform grids. 
For the 3-D simulations, the $Ra = 10^{7}$ cases also use non-uniform grids. 
Specifically, the mesh spacing in the wall-parallel directions remains uniform for all cases;
in the wall-normal direction, we adopt a cosine stretching function $y(\xi)=-\cos[\pi(\xi+1)/2]$ with $\xi \in [-1, 1]$  to cluster points \citep{bernardini2014velocity}. 
To validate the resolution, grid spacing in the wall-normal direction ${\Delta _y}$ and time interval ${\Delta _t}$ are compared against the Kolmogorov and Batchelor scales.
The Kolmogorov length scale is estimated by the global criterion  $\eta _K = \left( \nu_f ^3/\langle \varepsilon _u \rangle_{V,t} \right)^{1/4}$, the Batchelor length scale is estimated by  $\eta _B = \eta_K Pr^{-1/2}$ \citep{batchelor1959small,silano2010numerical} and the Kolmogorov time scale is estimated by $\tau_{\eta _K} = \sqrt {\nu_f /\langle \varepsilon _u \rangle_{V,t}} $. 
Here, $\varepsilon _u$ denotes the turbulent kinetic energy dissipation rates, and $\langle  \cdots \rangle _{V,t}$ denotes the volume- and time-averaged values. 
As shown in table \ref{tab:resolution}, the grid spacing in the wall-normal direction meets the criterion $\max ({\Delta _y}/{\eta _K},{\Delta _y}/{\eta _B}) \approx 1$, ensuring sufficient spatial resolution.
Additionally, the time intervals satisfy  $\max({\Delta _t} /\tau _{{\eta _K}}) \approx 0.007$, guaranteeing adequate temporal resolution.

We have varied both the diameter and the density of the particles in our simulations. 
Specifically,  we explored the parameter space of 5  $\mu$m  $ \leq {d_p} \leq $ 80 $\mu$m and 600 kg/m$^3$  $ \leq {\rho _p} \leq $  3000 kg/m$^3$.
Initially, the 2-D cell contains $N_{0}=100\times 50=5000$ particles, and the 3-D cell contains $N_{0}=150\times 50 \times 75=562 500$ particles.
Motivated by studies of sand grains and pollutants dispersed in the air ($\rho_{p}/\rho_{f} \gg 1$), we have the buoyancy ratio $\Lambda \approx 1/(\beta \Delta_{T})$ for all the simulation cases.
In our simulations, we fixed the temperature difference  $\Delta_{T}=5$ K, leading to  $\Lambda \approx 59.5$. 
Then, (\ref{eq:particle}) implies that we have one non-dimensional parameter, the Stokes number, to control the particle dynamics. 
The Stokes number based on the free-fall time scale is defined as $S{t_f} = {\tau _p}/{t_f} = {\rho _p}d_p^2/\left( {18{\mu _f}} \right)/\sqrt {H/(g\beta {\Delta _T})}$, and $St_{f}$ characterizes the large-scale effects. 
On the other hand, the Stokes number based on the Kolmogorov time scale is defined as  $S{t_K} = {\tau _p}/{\tau _{{\eta _K}} } = {\rho _p}d_p^2/\left( {18{\mu _f}} \right)/\sqrt {\nu_f /{{\left\langle {{\varepsilon _u}} \right\rangle }_{V,t}}}$, and $St_{K}$ characterizes the small-scale effects. 
For clarity in the following discussion, $St$ will denote $St_K$ to reflect the particle inertia relative to that of small-scale fluid motion  unless otherwise mentioned. 
We will use particle density ${\rho _p}$ and particle diameter ${d_p}$ to distinguish between different simulation cases.
Dimensional numerical values for various particle quantities of interest in terms of $St_f$ and $St_K$ are tabulated in the supplementary table available at \url{https://doi.org/10.1017/jfm.2024.936}.
In addition, we estimate the Kolmogorov length scale as  ${\eta _K} \geq $ 2.0 mm (see table \ref{tab:resolution}), which is around  25 times larger than the largest particle. 
The largest particle volume fraction of all cases is $\phi \sim \mathcal{O}(10^{-5})$, thus justifying the use of a one-way coupling approach to model the motion of point particles in this work. 
Even for large $St$, where sedimentation dominates over convective or shear motions, the back reaction of particles on the fluid is generally less significant if $\eta_{K}/d_{p}\gtrsim 10$ \citep{gore1989effect}. 
In Appendix \ref{appA},  we further examine the spatial distribution of the time-averaged local Kolmogorov length scale to support that the point-particle model can be safely used.
However, we acknowledge that this simplification holds primarily in dilute regimes \citep{balachandar2010turbulent}; in more concentrated suspensions, the back reaction can become important and should be considered.

	\begin{table}
		\centering
		\begin{tabular}{ccccccccc}
			$Ra$ & $Re_w$ & $Ri$ & Effective grid number &  $(\Delta_y)_{\max}/\eta_K$ & $(\Delta_y)_{\max}/\eta_B$ & $(\Delta_t)_{\max}/\tau_{\eta_K}$ & $\eta_K(mm)$ \\ 
		
			$10^7$ (2-D) & 0 & $\infty$ & 369$\times$198         & 0.62 & 0.52 & 0.0022 & 3.4 \\ 
			& 500 & 56.34 & 369$\times$198                 & 0.62 & 0.52 & 0.0019 & 3.4 \\ 
			& 1000 & 14.08 & 369$\times$198                & 0.66 & 0.56 & 0.0024 & 3.2 \\ 
			& 1500 & 6.26 & 369$\times$198                 & 0.74 & 0.63 & 0.0024 & 2.8 \\ 
			& 2000 & 3.52 & 369$\times$198                 &           -  & - & - & - \\ 

			$10^8$ (2-D) & 0 & $\infty$ & 792$\times$396         & 0.83 & 0.70 & 0.0020 & 2.7 \\ 
			& 1000 & 140.85 & 792$\times$396               & 0.96 & 0.81 & 0.0027 & 2.3 \\ 
			& 2000 & 35.21 & 792$\times$396                & 0.97 & 0.82 & 0.0025 & 2.3 \\ 
			& 3000 & 15.65 & 792$\times$396                & 0.97 & 0.82 & 0.0024 & 2.3 \\ 
			& 4000 & 8.80 & 792$\times$396                 & 0.97 & 0.82 & 0.0026 & 2.3 \\ 
			& 5000 & 5.63 & 792$\times$396                 & 0.91 & 0.76 & 0.0022 & 2.5 \\ 
			& 6000 & 3.91 & 1008$\times$504                & 0.59 & 0.49 & 0.0010 & 3.0 \\ 
			& 7000 & 2.87 & 1008$\times$504                & 0.55 & 0.47 & 0.0009 & 3.2 \\ 
			& 8000 & 2.20 & 1008$\times$504                & - & - & - & - \\ 

			$10^9$ (2-D) & 0 & $\infty$ & 1584$\times$792        & 1.01 & 0.85 & 0.0013 & 2.4 \\ 
			& 1000 & 1408.45 & 1584$\times$792             & 1.00 & 0.84 & 0.0013 & 2.4 \\ 
			& 2000 & 352.11 & 1584$\times$792              & 1.01 & 0.85 & 0.0012 & 2.4 \\ 
			& 3000 & 156.49 & 1584$\times$792              & 0.99 & 0.84 & 0.0012 & 2.4 \\ 
			& 4000 & 88.03 & 1584$\times$792               & 1.01 & 0.85 & 0.0011 & 2.4 \\ 
			& 5000 & 56.34 & 1584$\times$792               & 1.01 & 0.85 & 0.0013 & 2.4 \\ 
			& 6000 & 39.12 & 1584$\times$792               & 0.95 & 0.80 & 0.0011 & 2.5 \\ 
			& 9000 & 17.39 & 1890$\times$945               & 1.23 & 1.03 & 0.0018 & 2.6 \\ 
			& 12000 & 9.78 & 1890$\times$1188              & 0.87 & 0.73 & 0.0007 & 2.9 \\ 

			$10^{7}$ (3-D) & 0 & $\infty$ & 144$\times$256$\times$128  & 1.10 & 0.93 & 0.0054 &  2.2 \\ 
			& 2000 & 3.52 & 144$\times$256$\times$128          & 1.15 & 0.97 & 0.0073 &  2.1 \\ 
			& 4000 & 0.88 & 144$\times$256$\times$128          & 1.22 & 1.04 & 0.0068 &  2.0 \\ 
			\bottomrule

		\end{tabular}
		\caption{An \emph{a posteriori} check of spatial and temporal resolutions of the simulations.
The columns from left to right indicate the following: 
Rayleigh number $Ra$, wall-shear Reynolds number $Re_w$, Richardson number $Ri$, the effective grid number (i.e. the product of spectral element number and polynomial order), the ratio of maximum grid spacing in the wall-normal direction over the Kolmogorov length scale, the ratio of maximum grid spacing in the wall-normal direction over the Batchelor length scale, the ratio of maximum time interval over the Kolmogorov time scale, the Kolmogorov length scale.
Note that for the 2-D simulations at $Ra=10^{7}$ with $Re_{w}=2000$, and at $Ra=10^{8}$ with $Re_{w}=8000$, the flow is in the laminar state.}
		\label{tab:resolution}
	\end{table}

\section{Results and discussion}
\label{sec:3 Results}

\subsection{Particle transport in the convection cell}
\begin{figure}
	\centerline{\includegraphics[width=0.99\textwidth]{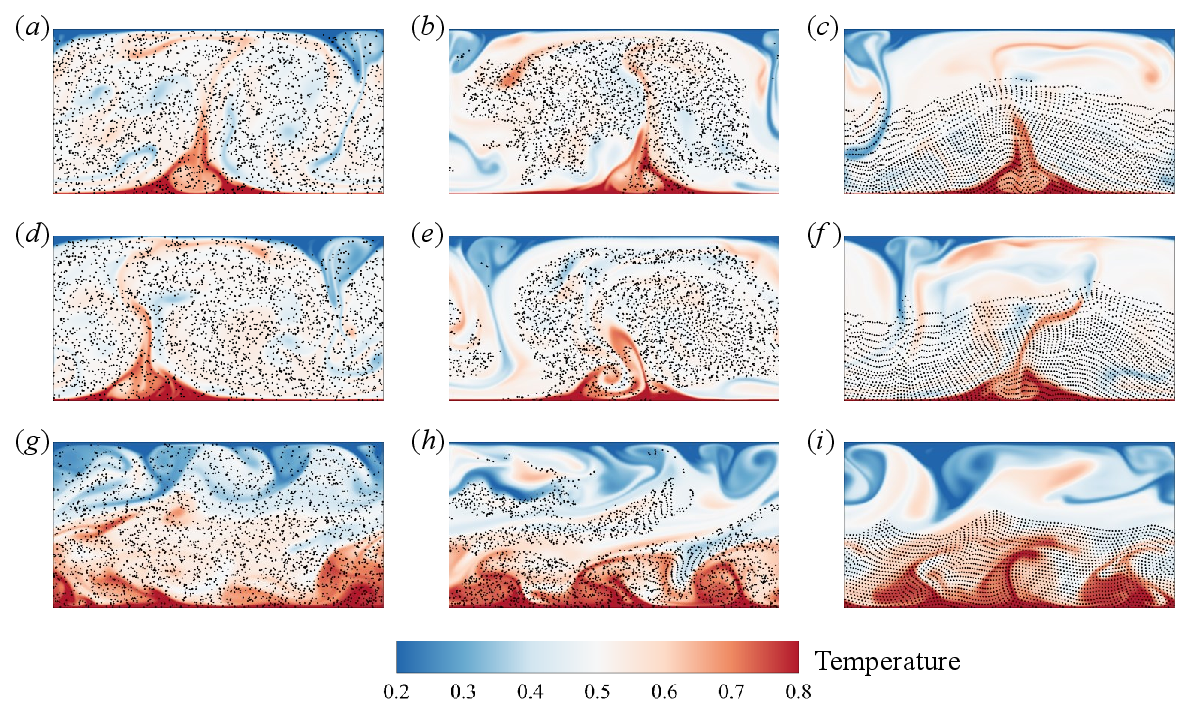}}
	\caption{Typical instantaneous temperature fields (contours) and particle positions (black dots, sizes artificially increased for better flow visualization)  from 2-D simulations at (\emph{a-c}) $Re_w$ = 0, (\emph{d-f}) $Re_w$ = 2000 and (\emph{g-i}) $Re_w$ = 6000, with $Ra=10^{8}$.
Particle density is fixed as ${\rho _p}= 3000$ kg/m$^3$, with diameters of  (\emph{a,d,g}) ${d_p} = 5$  $\mu$m,  (\emph{b,e,h}) ${d_p}=20$ $\mu$m and (\emph{c,f,i}) ${d_p} = 80$ $\mu$m.
Corresponding Stokes numbers are 
(\emph{a}) $St = 4.86 \times 10^{-4}$, (\emph{b}) $St = 7.78 \times 10^{-3}$, (\emph{c}) $St = 1.25 \times 10^{-1}$,
(\emph{d}) $St = 6.64 \times 10^{-4}$, (\emph{e}) $St = 1.06 \times 10^{-2}$, (\emph{f}) $St = 1.70 \times 10^{-1}$,
(\emph{g}) $St = 3.93 \times 10^{-4}$, (\emph{h}) $St = 6.29 \times 10^{-3}$ and (\emph{i}) $St = 1.01 \times 10^{-1}$. }
	\label{fig:temperature-2D}
\end{figure}

In figure \ref{fig:temperature-2D}, we show snapshots of temperature fields at various wall-shear Reynolds numbers $Re_w$ from 2-D simulations, and the corresponding movie can be viewed in supplementary movie 1. 
These snapshots are taken when half of the total particles remain suspended in the convection cell.
The canonical RB convection (see figures \ref{fig:temperature-2D}\emph{a-c}) consists of two large-scale rolls rotating in opposite directions. 
At lower $Re_w$ (see figures \ref{fig:temperature-2D}\emph{d-f}), the horizontal wall-shear effects are weak, and the dominant flow patterns are similar to those in the canonical RB convection. 
As the wall shear increases, the large-scale rolls undergo horizontal expansion, eventually leading to a transition into the zonal flow state (see figures \ref{fig:temperature-2D}\emph{g-i}). 
 Zonal flows are primarily horizontal flows typically aligned with the rotation axis in rotating systems \citep{goluskin2014convectively,zhang2020boundary,lin2021large}
or along a specific direction in non-rotating systems \citep{wang2020zonal,winchester2021zonal,jin2022shear}. 
These flows exhibit large-scale, coherent structures that can span the entire system and are often associated with banded structures in the flow field.
Previous 2-D studies on zonal flow were primarily conducted in convection cells with free-slip boundaries, which do not exert shear stress to slow down the fluid, and the periodicity of the flow system allows for a horizontal mean flow \citep{goluskin2014convectively,wang2020zonal,winchester2021zonal}. 
In contrast, our work shows that moving boundaries induce a horizontal mean flow, forming the zonal flow.
The overall trend in the 2-D flow pattern is similar to that in convection cells with rough shearing walls \citep{jin2022shear}.

Regarding particle movement,  the combined effects of particle diameter and particle density are reflected in the dimensionless Stokes number. 
For clarity of presentation, most results in this paper are shown at a fixed particle density but with varying particle diameters to present small, medium and large Stokes number cases.
Particles with small $St$ (see figures \ref{fig:temperature-2D}\emph{a,d,g}) are randomly distributed throughout the convection cell. 
These particles quickly adapt to changes in the surrounding flow, with little to no aggregation. 
For medium $St$ (see figures \ref{fig:temperature-2D}\emph{b,e,h}), we observe a greater concentration of particles over upwelling areas, with downwelling regions displaying fewer particles. 
This counterintuitive observation aligns with our previous findings \citep{xu2020transport} and independent research by \citet{patovcka2020settling,patovcka2022residence}. 
The physical interpretation is that, for particles circulating within large-scale rolls in the convection cell, when they approach the bottom boundary beneath downwellings, they are likely to be carried by the fast-moving edges of the rolls toward regions beneath upwellings. 
In these upwelling regions, particles may enter areas with slower flow due to small-scale irregularities caused by new plume generation. 
These low-velocity regions increase the likelihood of particle sedimentation, leading to an accumulation of particles beneath the upwellings.
In sheared convection cells, horizontal shear forces facilitate particle lateral dispersion, which is particularly apparent in figure \ref{fig:temperature-2D}(\emph{h}). 
For large $St$ (see figures \ref{fig:temperature-2D}\emph{c,f,i}), the inertia of the particles is dominant, causing them to settle and become nearly unresponsive to the carrier flow, similar to behaviour observed in quiescent fluid environments. 
As we will demonstrate later, even these large $St$ particles are slightly influenced by the turbulent flow.

\begin{figure}
	\centerline{\includegraphics[width=0.99\textwidth]{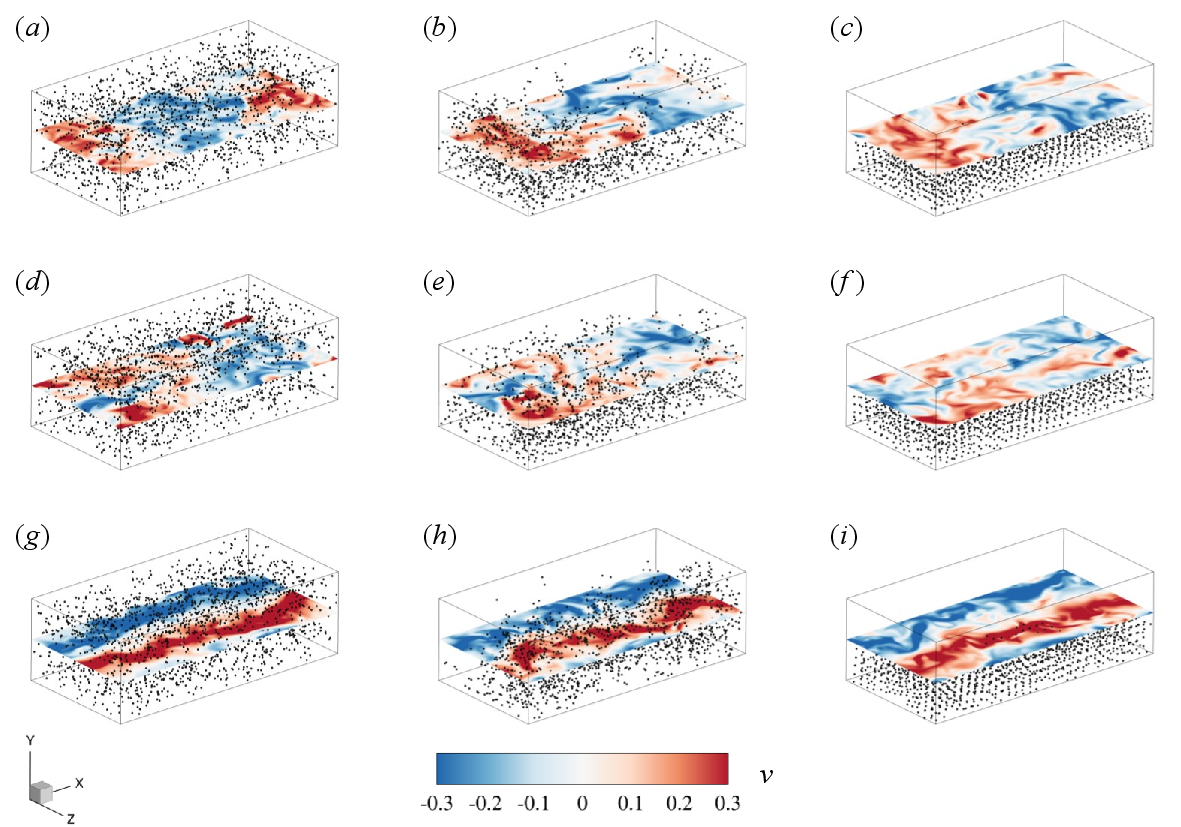}}
	\caption{ Typical instantaneous slice of velocity component $v$ at the channel centre plane (contours) and particle positions (black dots, sizes artificially increased for better flow visualization, here 0.8 \% of suspended particles are shown for clarity) from 3-D simulations at (\emph{a-c}) $Re_w$ = 0, (\emph{d-f}) $Re_w$ = 2000 and (\emph{g-i}) $Re_w$ = 4000, with $Ra=10^{7}$.
Particle density is fixed as ${\rho _p}= 3000$ kg/m$^3$, with diameters of (\emph{a,d,g}) ${d_p} = 5$  $\mu$m, (\emph{b,e,h}) ${d_p}=20$ $\mu$m and (\emph{c,f,i}) ${d_p} = 80$ $\mu$m.
Corresponding Stokes numbers are 
(\emph{a}) $St = 7.05 \times 10^{-4}$, (\emph{b}) $St = 1.13 \times 10^{-2}$, (\emph{c}) $St = 1.80 \times 10^{-1}$,
(\emph{d}) $St = 7.70 \times 10^{-4}$, (\emph{e}) $St = 1.23 \times 10^{-2}$, (\emph{f}) $St = 1.97 \times 10^{-1}$,
(\emph{g}) $St = 8.83 \times 10^{-4}$, (\emph{h}) $St = 1.41 \times 10^{-2}$ and (\emph{i}) $St = 2.26 \times 10^{-1}$. }
	\label{fig:velocity-3D}
\end{figure}

\begin{figure}
	\centerline{\includegraphics[width=0.99\textwidth]{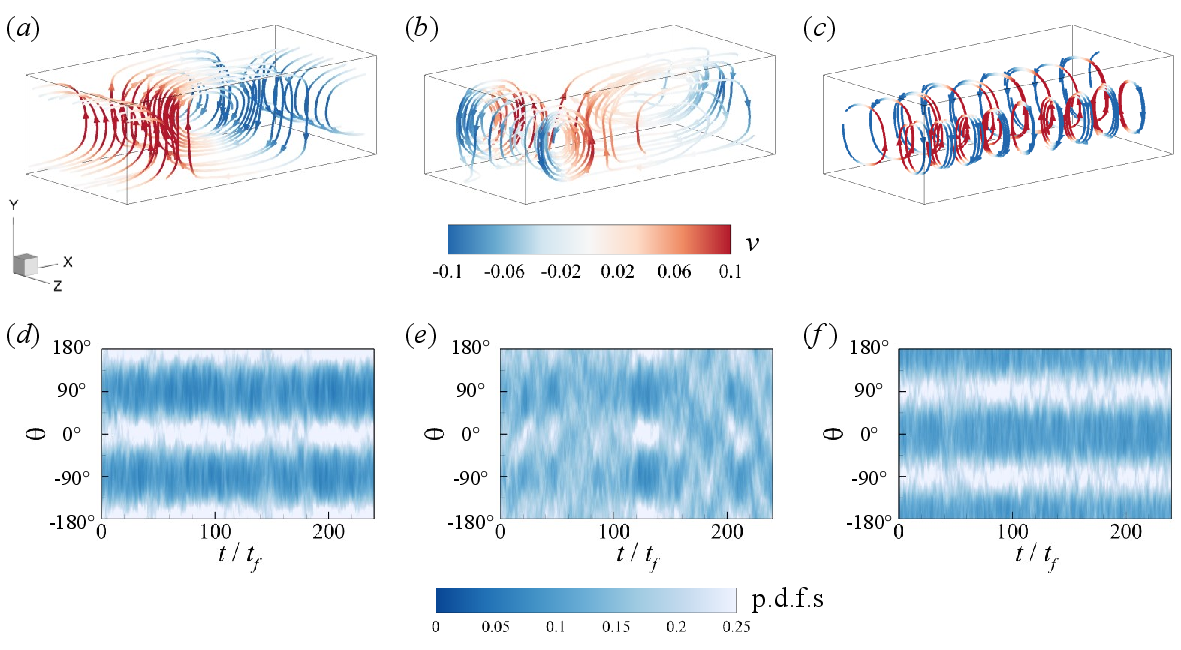}}
	\caption{ (\emph{a}-\emph{c}) Time-averaged streamlines coloured by the vertical velocity $v$, 
(\emph{d}-\emph{f}) time evolution of the p.d.f.s of the instantaneous velocity orientation $\theta$ 
at (\emph{a},\emph{d}) $Re_w$ = 0, (\emph{b},\emph{e}) $Re_w$ = 2000 and (\emph{c},\emph{f}) $Re_w$ = 4000, with $Ra=10^{7}$. }
	\label{fig:LSC-orientation}
\end{figure}

In figure \ref{fig:velocity-3D},  we further show snapshots of the velocity component $v$ at the channel centre plane and particle positions from 3-D simulations, and the corresponding movie can be viewed in supplementary movie 2. 
In thermal convection, rising fluids are warmer and falling fluids are colder.
At $Re_{w}=0$ and $Ra=10^{7}$, the flow structure resembles the canonical RB convection \citep{stevens2018turbulent}.
As $Re_{w}$ increases to 2000, the flow enters a transitional regime where the interaction between the imposed shear and buoyancy becomes significant.
Large, elongated thermal plumes start to transform into thin, straight, elongated streaks aligned in the streamwise direction. 
Eventually, streamwise-oriented rolls are formed at $Re_{w}=4000$ \citep{blass2020flow,pirozzoli2017mixed}.
To further elucidate the nature of these large-scale rolls, we present time-averaged streamlines coloured by the vertical velocity $v$, as shown in figures \ref{fig:LSC-orientation}(\emph{a}-\emph{c}).
At $Re_{w} = 0$, the averaged streamlines show a well-defined roll structure aligned in the spanwise direction. 
As $Re_{w}$ increases to 4000, the averaged streamlines transition to an organized pattern aligned with the streamwise direction.
Additionally, we measure the velocity orientation $\theta$, defined as the angle between the projection of the velocity vector onto the $x-z$ plane and the positive $x$-axis. 
We calculate the probability density functions (p.d.f.s) of $\theta$ for fluid nodes in the region $0.1H \le y \le 0.9H$, and present their time evolution in figures \ref{fig:LSC-orientation}(\emph{d}-\emph{f}). 
At $Re_{w}=0$, the p.d.f.s indicate high probability values consistently around  $0^\circ$ and $180^\circ$, suggesting that the roll axes are predominantly aligned in the spanwise direction, without exhibiting complex motions as seen in an RB cell with a larger length-to-width aspect ratio \citep{vogt2018jump,li2022counter,teimurazov2023oscillatory}. 
At $Re_{w}=4000$, the p.d.f.s shift to show high probability values around $\pm 90^\circ$, indicating that the roll axes are consistently aligned in the streamwise direction.

\begin{figure}
	\centerline{\includegraphics[width=0.99\textwidth]{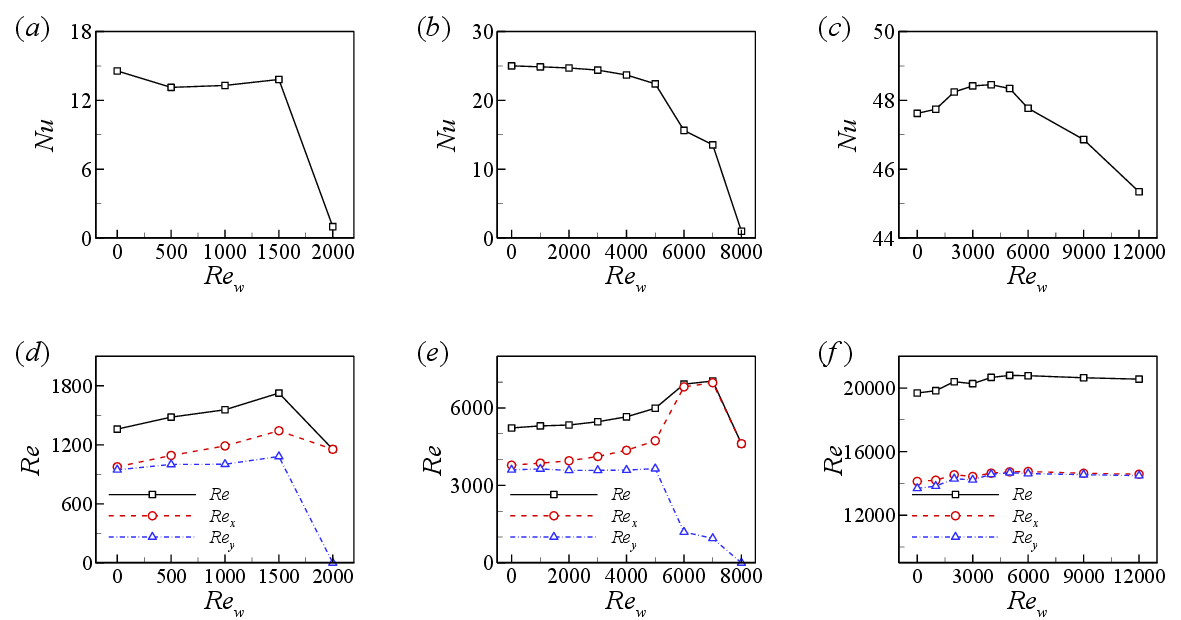}}
	\caption{ (\emph{a}-\emph{c}) Nusselt number and (\emph{d}-\emph{f}) Reynolds number as a function of $Re_w$:
(\emph{a},\emph{d}) with $Ra=10^{7}$, (\emph{b},\emph{e}) with $Ra=10^{8}$, (\emph{c},\emph{f}) with $Ra=10^{9}$.}
	\label{fig:ReNu}
\end{figure}

We then examine the global response parameters of the Nusselt number ($Nu$) and Reynolds number ($Re$) on the control parameter $Re_w$ in the  2-D flow system. 
Here, the heat transfer efficiency is calculated as  $Nu = 1 + \sqrt {PrRa} {\langle v^*T^*\rangle _{V,t}}$, and the global flow strength of the convection is calculated as  $Re = \sqrt {{{\langle \|\mathbf{u} \|^{2} \rangle }_{V,t}}} H/\nu_f $. 
Moreover, we consider the lateral and vertical components of $Re$, respectively, as $Re_{x} = \sqrt {{{\langle {u^2}\rangle }_{V,t}}} H/\nu_f$, $Re_{y} = \sqrt {{{\langle {v^2}\rangle }_{V,t}}} H/\nu_f $. 
Both $Nu$ and $Re$ exhibit sharp changes and abrupt transitions at $Re_w \simeq 1500$ with $Ra=10^7$ (see figures \ref{fig:ReNu}\emph{a},\emph{d}), at $Re_w \simeq 5000$ with $Ra=10^8$ (see figures \ref{fig:ReNu}\emph{b},\emph{e}),
suggesting a threshold for the competition between wall shear and buoyancy that leads to changes in the flow state from LSCs to zonal flows in 2-D simulations.
The heat transport behaviour in wall-sheared thermal turbulence is governed by the interaction between the buoyancy-driven LSC and the shear introduced by the moving walls. 
At higher wall shear, strong wall instabilities disrupt the large-scale rolls, which are crucial for carrying thermal energy away from the wall and efficiently transporting heat. 
When these large-scale structures are disrupted, small-scale turbulent motions become more prevalent, and the convective motions near the wall increase.
However, these motions become more localized and less effective in transporting heat across larger vertical distances, thus decreasing the overall efficiency of heat transport.
Regarding the velocity fields, LSCs exhibit both significant vertical and horizontal velocity components, while zonal flows in two dimensions show strong horizontal velocities with a lower vertical component.
At a fixed Rayleigh number, when the wall shear becomes sufficiently strong, the flow transitions to laminar Couette flow, causing the vertical flow velocity component to become nearly zero. 
For a Rayleigh number of $10^{7}$, this transition occurs approximately at $Re_{w} \simeq 2000$ (see figure \ref{fig:ReNu}\emph{d});
for a Rayleigh number of $10^{8}$, the transition happens at around $Re_{w} \simeq 8000$ (see figure \ref{fig:ReNu}\emph{e}). 
At an even higher Rayleigh number of $10^{9}$, based on the results of \citet{jin2022shear} and \citet{xu2023wall},  we speculate that the transitions to laminar Couette flow occurs at a much larger $Re_{w}$, beyond the parameter range explored in the present work.

\begin{figure}
	\centerline{\includegraphics[width=0.99\textwidth]{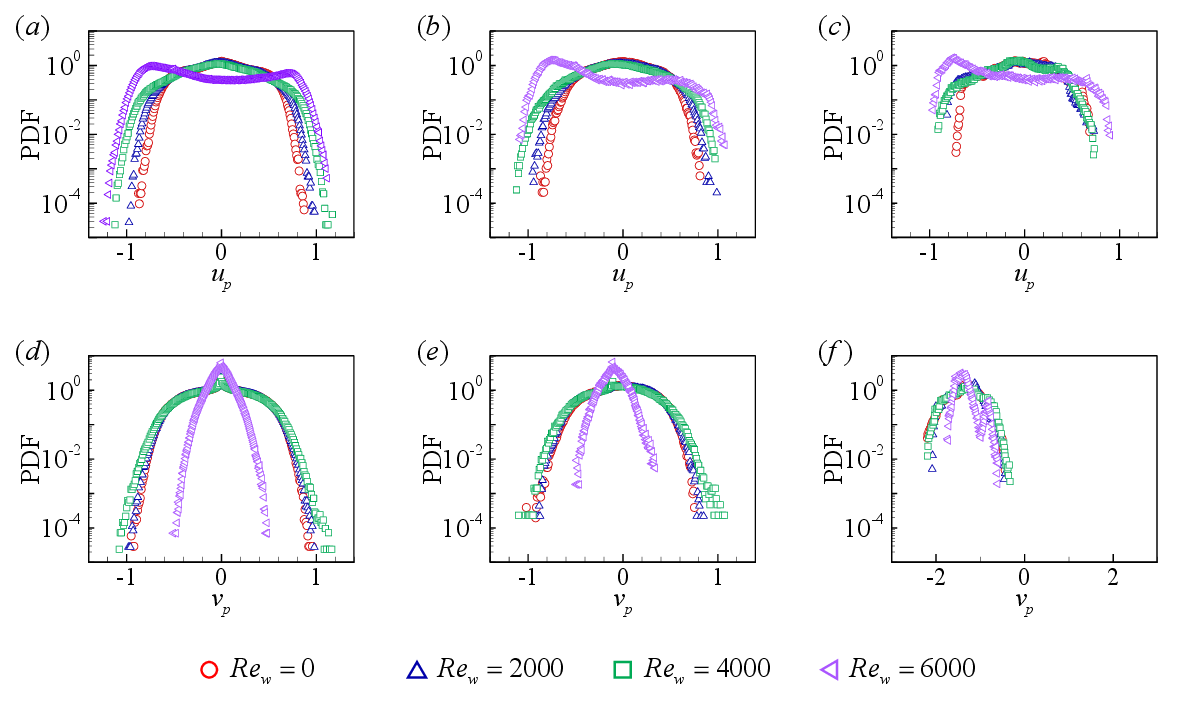}}
	\caption{The p.d.f.s of the particle velocity component (\emph{a-c}) in the horizontal direction ($u_p$), and (\emph{d-f}) in the vertical direction ($v_p$) with $Ra=10^{8}$. 
Particle density is fixed as ${\rho _p}= 3000$ kg/m$^3$, with diameters of (\emph{a,d}) ${d_p} = 5$  $\mu$m, (\emph{b,e}) ${d_p}=20$ $\mu$m and (\emph{c,f}) ${d_p} = 80$ $\mu$m.}
	\label{fig:UpVpPDF}
\end{figure}

We quantitatively describe the influence of wall shear on particle motion by analysing the p.d.f.s of their velocity components ($u_p$ and $v_p$), as shown in figure \ref{fig:UpVpPDF}. 
For particles with small and medium $St$ (see figures \ref{fig:UpVpPDF}\emph{a,b}), the p.d.f.s of $u_p$ are characterized by a single peak centred around zero, exhibiting symmetry for $Re_w \lesssim 5000$, suggesting a substantial likelihood of particles circulating within the LSC of convective flows. 
As $Re_w$ increases to 6000, the p.d.f.s of $u_p$ change to a bimodal distribution with two peaks off zero value,  which can be attributed to the horizontal drift of particles caused by the wall shear. 
In addition, with increasing $St$, the peak corresponding to the negative $u_p$ exceeds that of the positive, suggesting an increased likelihood for particle suspension in the lower domain of the cell due to sedimentation (note the bottom wall is moving in the negative $x$-direction). 
For large $St$ (see figure \ref{fig:UpVpPDF}\emph{c}), we can deduce that these particles are still minorly influenced by the turbulent flow because they exhibit horizontal motions with finite positive or negative $u_p$. 
Regarding the p.d.f.s of $v_p$, for small and medium $St$ (see figures \ref{fig:UpVpPDF}\emph{d,e}), their shapes are much narrower at $Re_w$ = 6000 compared with those at lower $Re_w$, indicating attenuated vertical motions in the zonal flow. 
This trend becomes more pronounced for large $St$ (see figure \ref{fig:UpVpPDF}\emph{f}), where $v_p$ values are predominantly negative.
This implies vertical updrafts are insufficient to counter the gravitational settling, preventing these particles from rising with the flow currents.

\begin{figure}
	\centerline{\includegraphics[width=0.99\textwidth]{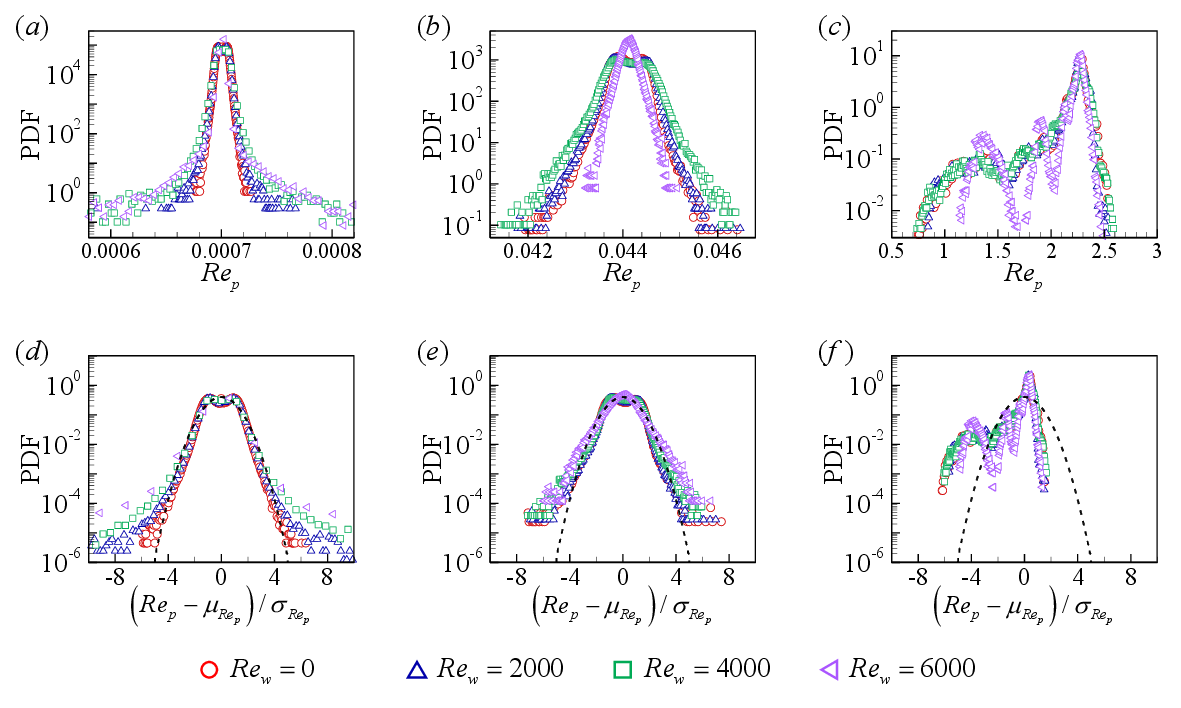}}
	\caption{(\emph{a-c}) The p.d.f.s of particle Reynolds number $Re_p$ and (\emph{d-f}) the p.d.f.s of normalized particle Reynolds number  $(Re_{p} - {\mu _{Re_{p}}})/{\sigma _{Re_{p}}}$  with $Ra=10^{8}$. 
Particle density is fixed as ${\rho _p}= 3000$ kg/m$^3$, with diameters of (\emph{a,d}) ${d_p} = 5$  $\mu$m, (\emph{b,e}) ${d_p}=20$ $\mu$m and (\emph{c,f}) ${d_p} = 80$ $\mu$m. 
Dashed lines in (\emph{d-f}) represent the Gaussian distribution.}
	\label{fig:RepPDF}
\end{figure}

We examine the p.d.f.s of the particle Reynolds number $Re_{p} = {d_p}|{{\boldsymbol{u}}_f} - {{\boldsymbol{u}}_p}|/{\nu_f}$, as shown in figure \ref{fig:RepPDF}. 
We can see from figures \ref{fig:RepPDF}($a,b$) that, for particles with small $St$, $Re_p$ is generally less than $10^{-3}$; 
for medium $St$, $Re_p$ is generally less than $10^{-1}$. 
In the presence of a pronounced LSC within the flow, the p.d.f.s of $Re_p$ manifest dual peaks, corresponding to the particles ascending or descending with the LSC. 
As wall shear increases, the $Re_p$ will also increase and exhibit enhanced intermittency. 
However, when $Re_w$ further increases up to 6000, $Re_p$ decreases and the p.d.f.s of $Re_{p}$ change from a bimodal to unimodal distribution. 
This shift is attributed to the decreased vertical velocity component $v_p$ of the particles within the convective flow, as evidenced by figures \ref{fig:UpVpPDF}(\emph{d-f}). 
For large $St$, $Re_p$ is as large as $Re_p\sim O(1)$ and spans a more extensive range, as shown in figure \ref{fig:RepPDF}(\emph{c}), which emphasizes the necessity of incorporating a drag correction factor $f(Re_{p})$ into the drag force calculations. 
We further plot the p.d.f.s of normalized particle Reynolds number  $(Re_{p} - {\mu _{Re_{p}}})/{\sigma _{Re_{p}}}$, as shown in figures \ref{fig:RepPDF}(\emph{d-f}). 
For particles with small $St$ (see figure \ref{fig:RepPDF}\emph{d}), the p.d.f.s of $Re_p$ are close to the Gaussian distribution (denoted by the dashed line in the figure). 
However, with an increase in $Re_w$, the deviations in the tails of the p.d.f.s. become more pronounced, indicating more intermittency events due to the influence of wall shear. 
For medium $St$ (see figure \ref{fig:RepPDF}\emph{e}), the p.d.f.s generally follow the Gaussian distribution, although there appear to be deviations with heavier tails, which implies there are more instances of $Re_p$ values deviating from the mean. 
For large $St$ (see figure \ref{fig:RepPDF}\emph{f}), the p.d.f.s of $Re_p$ are distinctly different from a Gaussian distribution, indicating a distinct distribution pattern where particle motion is nearly unaffected by the turbulence.

\begin{figure}
	\centerline{\includegraphics[width=0.99\textwidth]{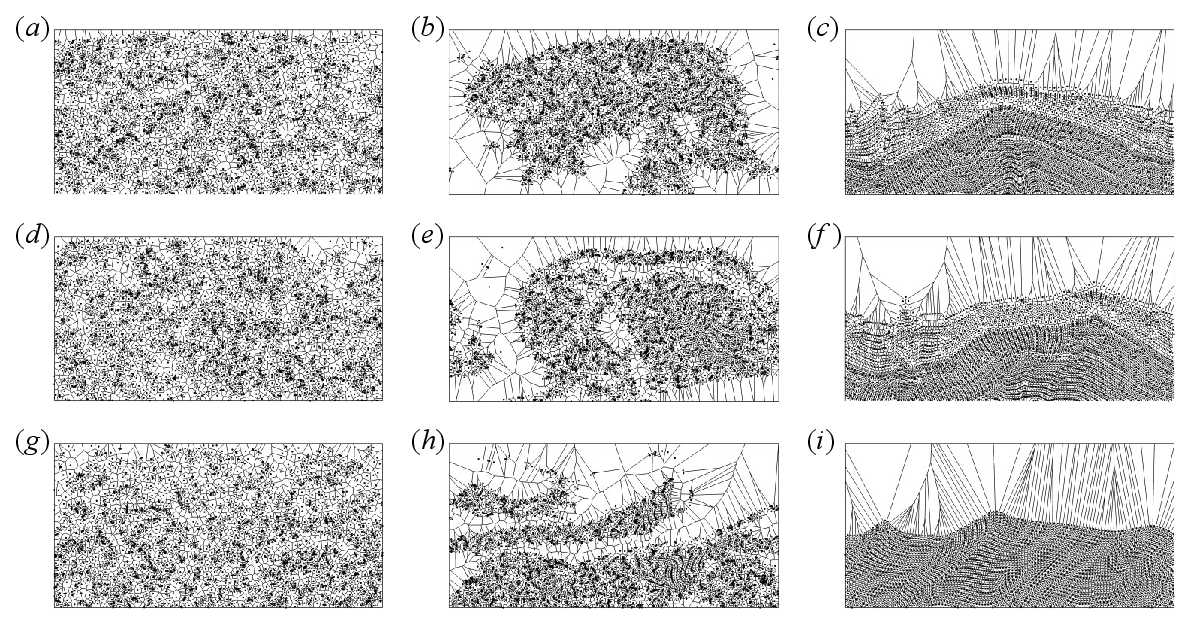}}
	\caption{Typical instantaneous Vorono\"{i} diagrams for particles at (\emph{a-c}) $Re_w = 0$, (\emph{d-f}) $Re_w = 2000$, (\emph{g-i}) $Re_w = 6000$ with $Ra=10^{8}$.
Particle density is fixed as ${\rho _p}= 3000$ kg/m$^3$, with diameters of (\emph{a,d,g}) ${d_p} = 5$  $\mu$m, (\emph{b,e,h}) ${d_p}=20$ $\mu$m and (\emph{c,f,i}) ${d_p} = 80$ $\mu$m.}
	\label{fig:voronoi}
\end{figure}

Previous studies have shown that, even in homogeneous isotropic turbulence, the particles may not distribute homogeneously but exhibit  clustering behaviour \citep{wang1993settling,bosse2006small,calzavarini2008dimensionality}. 
To identify the clustering behaviour of particles during their transport in the wall-sheared convection cell, we employ a quantitative analysis using Vorono\"{i} diagrams \citep{monchaux2010preferential,monchaux2012analyzing}. 
A Vorono\"{i} cell is defined as a collection of points closer to its corresponding particle than to any other; thus, the area (or volume) of a Vorono\"{i} cell inversely correlates with local particle concentration. 
In figure \ref{fig:voronoi}, we show Vorono\"{i} diagrams for the particle distribution, corresponding to the instantaneous state presented in figure \ref{fig:temperature-2D}. 
For particles with small $St$ (see figures \ref{fig:voronoi}\emph{a,d,g}),  Vorono\"{i} cells are of similar size and relatively evenly distributed in space, indicating a low degree of aggregation.
For medium $St$ (see figures \ref{fig:voronoi}\emph{b,e,h}), the areas of their Vorono\"{i} polygon exhibit much higher variability, indicating an increased level of spatial inhomogeneity. 
For large $St$ (see figures \ref{fig:voronoi}\emph{c,f,i}), large voids are formed at the upper part of the convection cell, containing few to no particles. 
This is because most particles settle down and do not circulate in the cell, as revealed by the predominantly negative  $v_p$ velocity components shown in figures \ref{fig:UpVpPDF}(\emph{f}).

\begin{figure}
	\centerline{\includegraphics[width=0.99\textwidth]{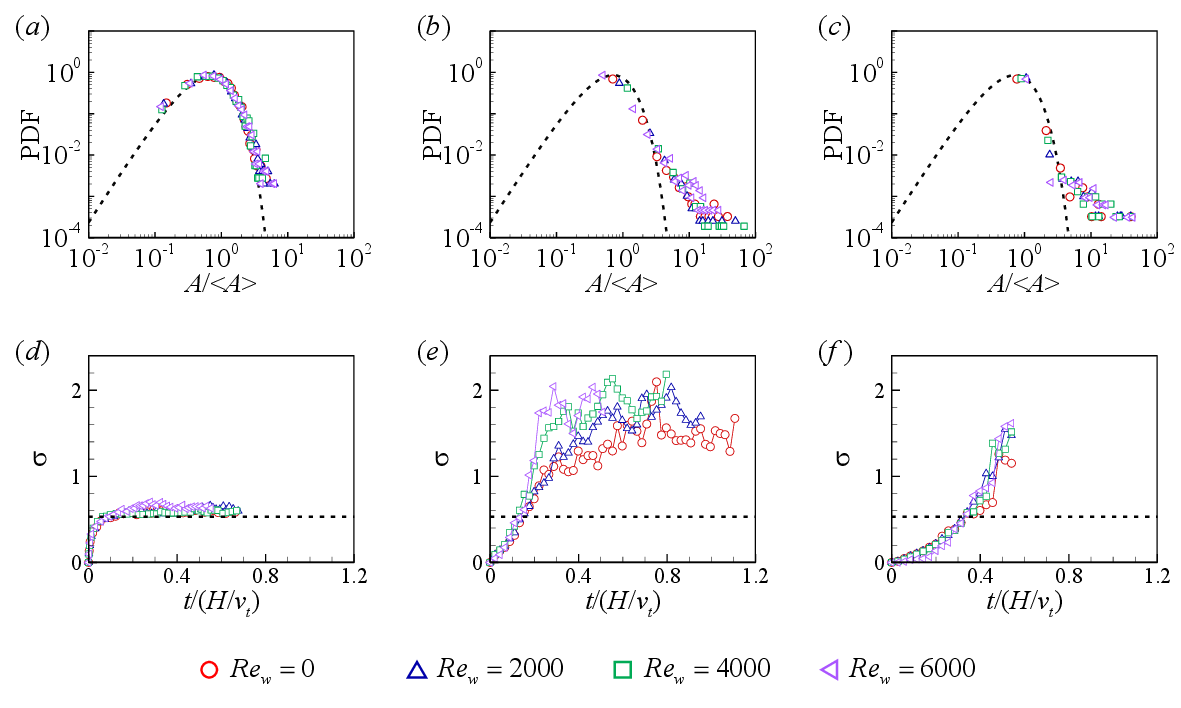}}
	\caption{(\emph{a-c}) The p.d.f.s of the normalized Vorono\"{i} cell area  $A/\langle A \rangle$ corresponding to the instantaneous state presented in figure \ref{fig:voronoi}, and (\emph{d-f}) time evolution of the standard deviation of the normalized Vorono\"{i} cell area. 
Panels show (\emph{a,d}) ${d_p} = 5$  $\mu$m, (\emph{b,e}) ${d_p}=20$ $\mu$m and (\emph{c,f}) ${d_p} = 80$ $\mu$m.}
	\label{fig:voronoiPDF}
\end{figure}

We then analyse the statistical distribution of the Vorono\"{i} cell area $A$. 
Figure \ref{fig:voronoiPDF} shows the p.d.f.s of the normalized Vorono\"{i} cell area  $A/\left\langle A \right\rangle $, corresponding to the instantaneous state presented in figure \ref{fig:voronoi}. 
Here, $\left\langle A \right\rangle$ represents the mean value of $A$ at the current moment.
For particles with small $St$ (see figure \ref{fig:voronoiPDF}\emph{a}), the Vorono\"{i} cells are randomly distributed throughout the flow field, and the distribution of their areas satisfies the $\Gamma$ distribution \citep{ferenc2007size,tagawa2012three}
\begin{equation}
f_d(m) = \frac{\left[(3d + 1)/2\right]^{(3d + 1)/2}}{\Gamma \left[(3d + 1)/2 \right] } m^{(3d - 1)/2} \exp \left( - \frac{3d + 1}{2} m \right), \quad m = \frac{A}{\langle A \rangle}
\end{equation}
where $d = 2$ denotes the analysis in two dimensions. 
For medium $St$ (see figure \ref{fig:voronoiPDF}\emph{b}), the p.d.f.s have a peak near unity, yet exhibit pronounced tails indicative of larger Vorono\"{i} cell areas, deviating from the $\Gamma$ distribution. 
This deviation from the $\Gamma$ distribution signals a departure from randomness, indicative of local preferential concentration. 
For large $St$ (see figure \ref{fig:voronoiPDF}\emph{c}), the p.d.f.s exhibit more pronounced tails towards larger Vorono\"{i} cell areas, indicating intensified void formation.

Figures \ref{fig:voronoiPDF}(\emph{d-f}) further shows the time evolution of the standard deviation  $\sigma $ of the normalized Vorono\"{i} cell area, serving as an indicator of spatial inhomogeneity. 
Due to particle deposition on the wall and constantly reducing particle numbers, the particles statistics cannot reach a statistically stationary state. 
Here, we only plot the time series from the initial state to the time at which half of the total particles remain suspended in the convection cell. 
For particles with small $St$ (see figure \ref{fig:voronoiPDF}\emph{d}), the standard deviation is the lowest, approaching the theoretical expectation for the $\Gamma$ distribution (i.e. $\sigma  = 0.53$, denoted by the dash-dotted line). 
For medium $St$ (see figure \ref{fig:voronoiPDF}\emph{e}), the greatest clustering is observed as the standard deviation of their Vorono\"{i} area significantly exceeds 0.53. 
For large $St$ (see figure \ref{fig:voronoiPDF}\emph{f}), the clustering also occurs; however, the influence of variations in wall-shear strength $Re_w$ appears to be negligible in contrast to those with medium $St$.

Based on the above analysis, we arrive at the following mathematical model for the spatial distribution of particle concentration. 
We denote the initial concentration of particles as   ${n_0} = {N_0}/(LHW)$, where $N_{0}$ is the initial number of particles laden in the fluid.
We define the instantaneous particle line density averaged over a horizontal cross-section as $n(\tau,y) = N(\tau,y)/(LW)$ at a given time $\tau$ and height $y$, where $N(\tau,y)$ is the instantaneous particle number at time $\tau$ and height $y$.
We also define the time-averaged local particle concentration over a horizontal cross-section and along the cell height as $\langle n(y) \rangle _{t}$, where the average $\langle \cdot \rangle_{t}$ is 
calculated over time $t$.
In transient processes, the particle concentration changes over time, and the time-averaged concentration represents the overall behaviour of the particles during a specified time interval, which is crucial for understanding the long-term behaviour of the system, even when the process is not in a steady state.

For particles with small $St$, they are randomly distributed throughout the convection cell, and  they exhibit an exponential decay rate as ${N(\tau)} = N_{0}\exp \left[ { -\tau}/\left( {H/{v_t}} \right) \right]$ \citep{martin1989fluid}, with $N(\tau)$ being the number of particles that remain laden in the convection cell at time $\tau$. 
Following the assumption of \citet{martin1989fluid},  turbulent convection ensures that particles are mixed homogeneously across the flow, which leads to $N(\tau,y)=N(\tau)/H$, and particle concentration is uniform across the horizontal plane of the fluid. 
Thus, at a given time $\tau$ and height $y$,  the instantaneous particle line density $n(\tau,y)$ is 
\begin{equation}
n(\tau,y) = \frac{N(\tau,y)}{LW} = \frac{N(\tau)}{LHW} 
= \frac{N_0 \exp \left( - \frac{\tau}{H/v_t} \right)}{LHW} 
= n_0 \exp \left( - \frac{\tau}{H/v_t} \right)
\end{equation}
Then, the horizontal- and time-averaged local particle concentration,  which represents the number density of particles suspended in the convection cell over a given time period and horizontal cross-section, for small $St$, is  
\begin{equation}
\frac{\left\langle n(y) \right\rangle_t}{n_0} 
= \frac{\left( \int_0^t n(\tau, y) \, d\tau \right) / t}{n_0} 
= - \frac{H}{v_t t} \left[ \exp \left( - \frac{t}{H / v_t} \right) - 1 \right]
\label{eq:smallStLocalConcentration}
\end{equation}
 In the above, $\left\langle n\left( y \right) \right\rangle_{t}$ is further normalized by initial concentration $n_0$.

For large $St$, sedimentation occurs almost independently of the carrier flow, analogous to their settling behaviour in quiescent fluid environments. 
Thus, we can ignore their lateral motion and simply assume they settle at a constant speed of terminal velocity  $v_t$. 
From the initial state of uniform distribution, after a duration of $\tau$, the particle ensemble settles a vertical distance of  $v_t \tau$. 
Thus,  the instantaneous particle line density $n(\tau, y)$ is
\begin{equation}
n(\tau, y) = 
\begin{cases}
0, & H - v_t \tau < y \le H \\
n_0, & y \le H - v_t \tau
\end{cases}
\end{equation}
We consider at the time $\tilde{t} \in \left[ {0,t} \right]$ the instantaneous local particle concentration $n(\tilde{t}, y)$ that decreases from  ${n_0}$ to 0 across a height interval  $\tilde y$, where  $\tilde y \in \left[ {H - {v_t}\tilde{t},H} \right]$, resulting in  $H - {v_t}\tilde t = \tilde y$. 
Thus, the horizontal- and time-averaged local particle concentration (further normalized by  ${n_0}$),  for large $St$, is  
\begin{equation}
\frac{\left\langle n(y) \right\rangle_t}{n_0} 
= \frac{\left( \int_0^t n(\tau, y) \, d\tau \right) / t}{n_0} 
= \begin{cases}
\frac{\left( \int_0^{\tilde{t}} n_0 \, d\tau \right) / t}{n_0} + \frac{\left( \int_{\tilde{t}}^t 0 \, d\tau \right) / t}{n_0}, & H - v_t t \le y \le H \\[16pt]
\frac{\left( \int_0^t n_0 \, d\tau \right) / t}{n_0}, & y \le H - v_t t
\end{cases}
\end{equation}
The above equation can be further simplified to
\begin{equation}
\frac{\left\langle n(y) \right\rangle_t}{n_0} = 
\begin{cases}
\frac{H - y}{v_t t}, & H - v_t t \le y \le H \\
 \\
1, & y \le H - v_t t
\end{cases}
\label{eq:largeStLocalConcentration}
\end{equation}

We choose the time-average $\left\langle \cdot \right\rangle _{t_{1/2}}$ from the initial state up to the state when half of the total particles deposit on the wall (a duration referred to as half-life time $t_{1/2}$ from now on). 
For particles with small $St$,  they exhibit an exponential deposition rate \citep{martin1989fluid}. 
The half-life time can be obtained by setting $N_s/N_0 = 1 - \exp \left[ { - t_{1/2}/\left( {H/{v_t}} \right)} \right]=1/2$, which further gives $t_{1/2} = \left( {H\ln 2} \right)/{v_t}$. 
Here, $N_{s}$ is the number of particles that have settled and deposited on the wall.
Substituting this time duration into (\ref{eq:smallStLocalConcentration}), we derive the horizontal- and time-averaged local particle concentration for small $St$ as
\begin{equation}\label{eq:SSt_concen}
\frac{\left\langle n(y) \right\rangle_{t_{1/2}}}{n_0} = \frac{1}{2 \ln 2}
\end{equation}
For large $St$,  the deposition rate on the wall follows a linear law as derived from Stokes’ law. 
The half-life time can be obtained by setting $N_s/N_0 = t_{1/2}/(H/v_{t})=1/2$, which further gives
$t_{1/2} = 0.5H/{v_t}$. 
Substituting this into (\ref{eq:largeStLocalConcentration}), we derive the horizontal- and time-averaged local particle concentration for large $St$ as
\begin{equation}\label{eq:LSt_concen}
\frac{\left\langle n(y) \right\rangle_{t_{1/2}}}{n_0} = 
\begin{cases}
2\left( 1 - \frac{y}{H} \right), & 0.5H \le y \le H \\
 \\
1, & y \le 0.5H
\end{cases}
\end{equation}

\begin{figure}
	\centerline{\includegraphics[width=0.99\textwidth]{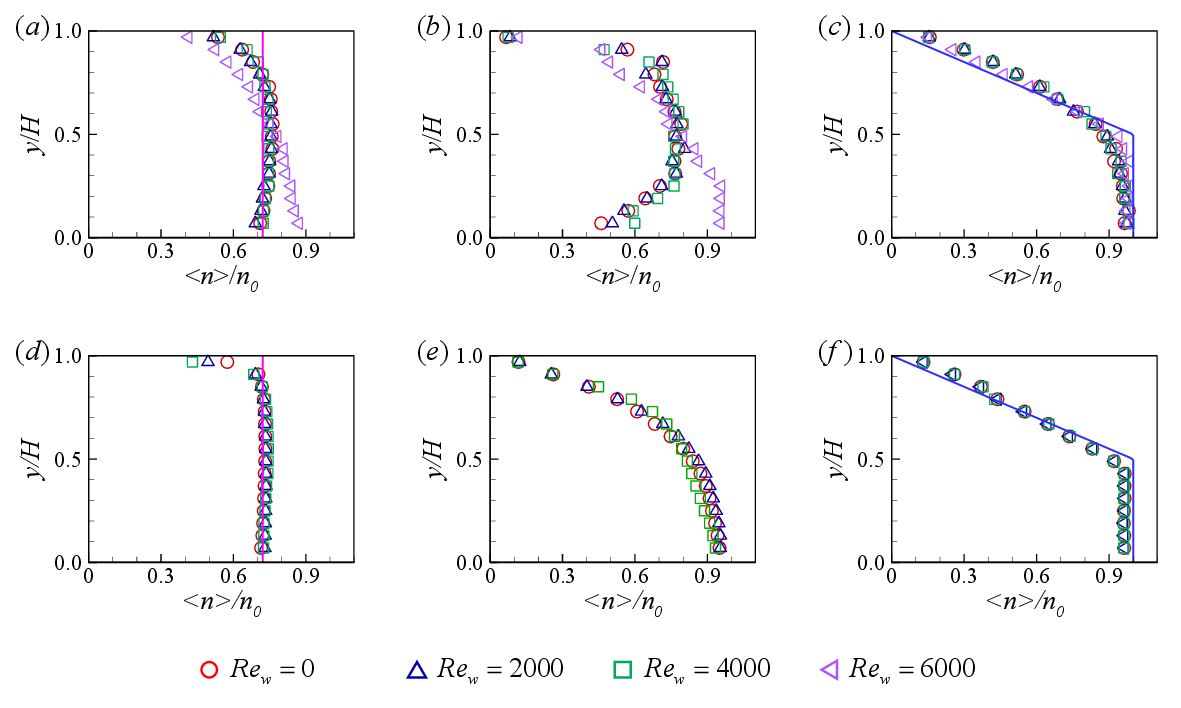}}
	\caption{Vertical profiles of the horizontal- and time-averaged local particle concentration  obtained from (\emph{a-c}) the 2-D simulation with $Ra=10^{8}$ and (\emph{d-f}) the 3-D simulation with $Ra=10^{7}$. 
Particle density is fixed as ${\rho _p}= 3000$ kg/m$^3$, with diameters of (\emph{a},\emph{d}) ${d_p} = 5\ \mu$m, (\emph{b},\emph{e}) ${d_p}=20\ \mu$m and (\emph{c},\emph{f}) ${d_p} = 80\ \mu$m.
The pink solid line in (\emph{a},\emph{d}) corresponds to (\ref{eq:SSt_concen}), and the blue solid line in (\emph{c},\emph{f}) corresponds to (\ref{eq:LSt_concen}).}
	\label{fig:numberDensity}
\end{figure}

In figure \ref{fig:numberDensity}, we compare the theoretical estimations of ${\left\langle {n\left( y \right)} \right\rangle _{t_{1/2}}}/{n_0}$, namely (\ref{eq:SSt_concen}) and (\ref{eq:LSt_concen}), with  both 2-D and 3-D simulation results. 
For particles with small $St$ (see figures \ref{fig:numberDensity}\emph{a},\emph{d}), the local particle concentration generally agrees with the theoretical predictions except for the discrepancy at $y > 0.8 H$. 
This discrepancy may be attributed to the finite $St$ effect (i.e. a small but not non-zero $St$), which results in a reduced particle presence in the upper boundary layer. 
At $Re_w = 6000$ in the 2-D simulations, the flow transitions from the LSC to zonal flow, thus the assumption of a spatially uniform particle concentration proposed by \citet{martin1989fluid} may no longer be valid, leading to significant deviation from (\ref{eq:SSt_concen}). 
Specifically, when the LSC transitions to a zonal flow state, the updrafts are significantly weakened.
This weakening results in fewer particles drifting upwards and more particles settling down, leading to a higher particle concentration at the bottom of the zonal flow compared with the LSC state.
For medium $St$ (see figures \ref{fig:numberDensity}\emph{b},\emph{e}), the absence of a corresponding model precludes a direct comparison; however, we provide a qualitative description of particle concentration. 
When the LSC is present, the local particle concentration appears to be consistent in the bulk region, but exhibits an increase with increasing $Re_w$ in the bottom boundary layer, implying that wall shear accelerates particle sedimentation. 
At $Re_w = 6000$ in the 2-D simulations, the changes in flow states lead to distinctly different profiles for the local particle concentration compared with those at lower $Re_w$. 
For large $St$ (see figures \ref{fig:numberDensity}\emph{c},\emph{f}), the local particle concentration generally agrees with the theoretical prediction (\ref{eq:LSt_concen}) across all $Re_w$, highlighting again that the particles' motion is nearly not affected by the turbulent flows of the carrier fluid (i.e. either LSC or zonal flow state). 
When the wall shear becomes sufficiently strong, the vertical flow velocity component weakens significantly. 
Since the settling velocity of particles is strongly influenced by the vertical flow velocity component, the settling velocity of particles then approaches that in a stationary flow field (i.e. the terminal velocity $v_{t}$). 
This explains why the concentration profile for $Re_{w} = 6000$ in figure \ref{fig:numberDensity}(\emph{b})  is similar to those in figure \ref{fig:numberDensity}(\emph{c})  for the 2-D simulation.

\subsection{Particle deposition on the wall}

\begin{figure}
 	\centerline{\includegraphics[height=0.52\textheight]{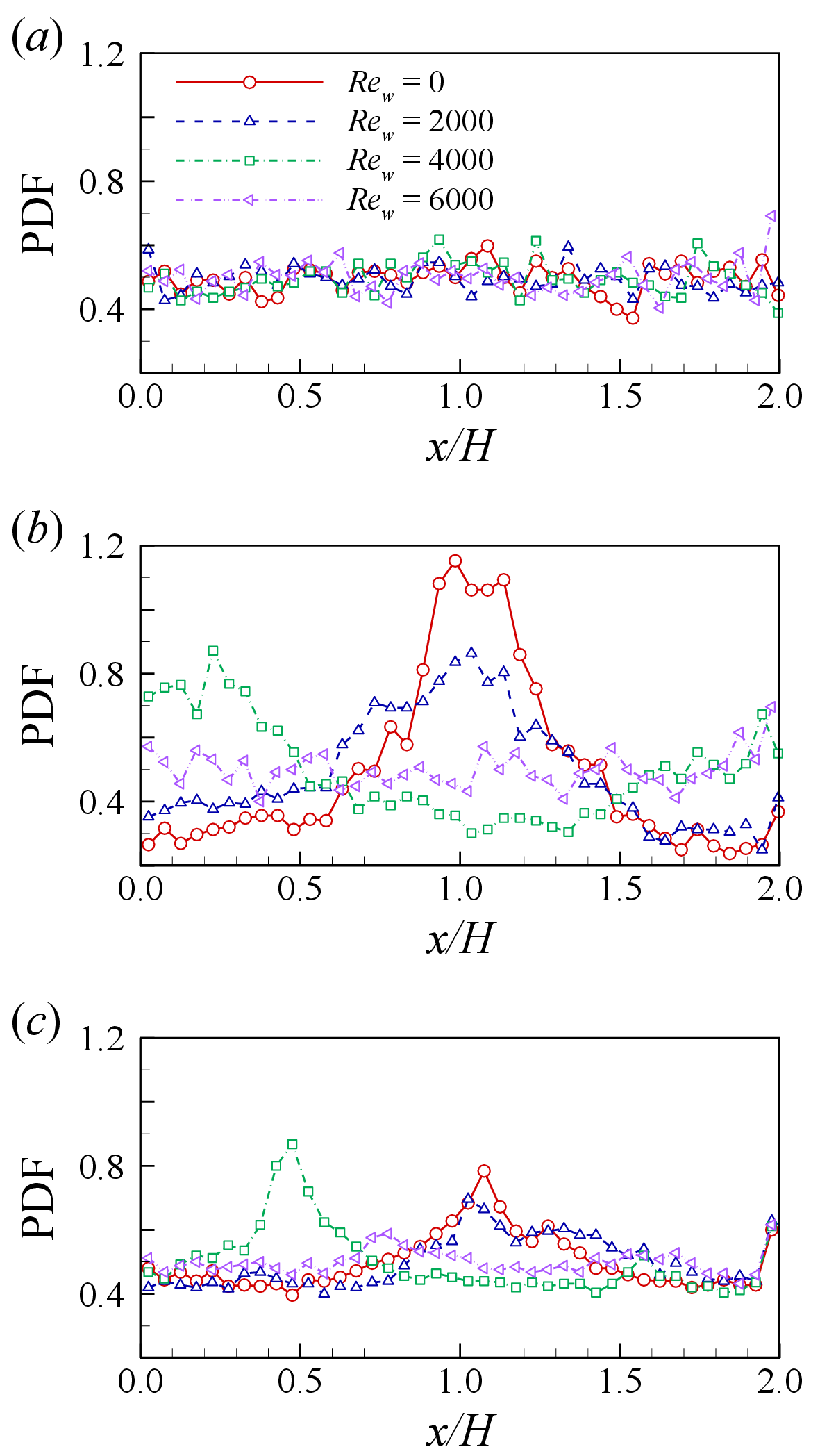}}
 	\caption{The p.d.f.s of particle deposition locations along the bottom wall  with $Ra=10^{8}$.  
Particle density is fixed as ${\rho _p}= 3000$ kg/m$^3$, with diameters of (\emph{a}) ${d_p} = 5\ \mu$m, (\emph{b}) ${d_p}=20\ \mu$m and (\emph{c}) ${d_p} = 80\ \mu$m.}
 	\label{fig:settlingdistributionpdf}
\end{figure}

To quantify the particle deposition behaviour, we first examine the p.d.f.s of their deposition locations along the bottom wall, as shown in figure \ref{fig:settlingdistributionpdf}. 
In our simulations, the particles adhere to the wall upon contact, and we then mark them as deposited and record the contact position as the deposition location. 
For particles with small $St$ (see figure \ref{fig:settlingdistributionpdf}\emph{a}), the deposition locations are uniformly distributed along the bottom wall, aligning with the assumption of a spatially uniform particle concentration. 
For medium $St$ (see figure \ref{fig:settlingdistributionpdf}\emph{b}), there is pronounced aggregation behaviour for deposition positions along the bottom wall. 
With the increase of $Re_w$, the degree of aggregation weakens, possibly because wall shear causes the boundary layers to move horizontally, thereby dispersing the particles to deposit more evenly along the bottom wall. 
We also observed a correlation between the regions of particle accumulation on the bottom wall and the rising of hot plumes (compared with figure \ref{fig:temperature-2D}), as revealed in our previous findings \citep{xu2020transport} and independent research by \citet{patovcka2020settling,patovcka2022residence}. 
The accumulation of heavy particles beneath upwellings seems somewhat counterintuitive, as one might expect these particles to accumulate primarily beneath significant downwellings. 
For large $St$ (see figure \ref{fig:settlingdistributionpdf}\emph{c}), the uniformity of the deposition pattern is restored, mainly attributed to the reduced influence of the carrier flow on these particles, resulting in an initial uniform distribution less affected by the carrier flow, unlike their low $St$ counterparts.

\begin{figure}
	\centerline{\includegraphics[width=0.99\textwidth]{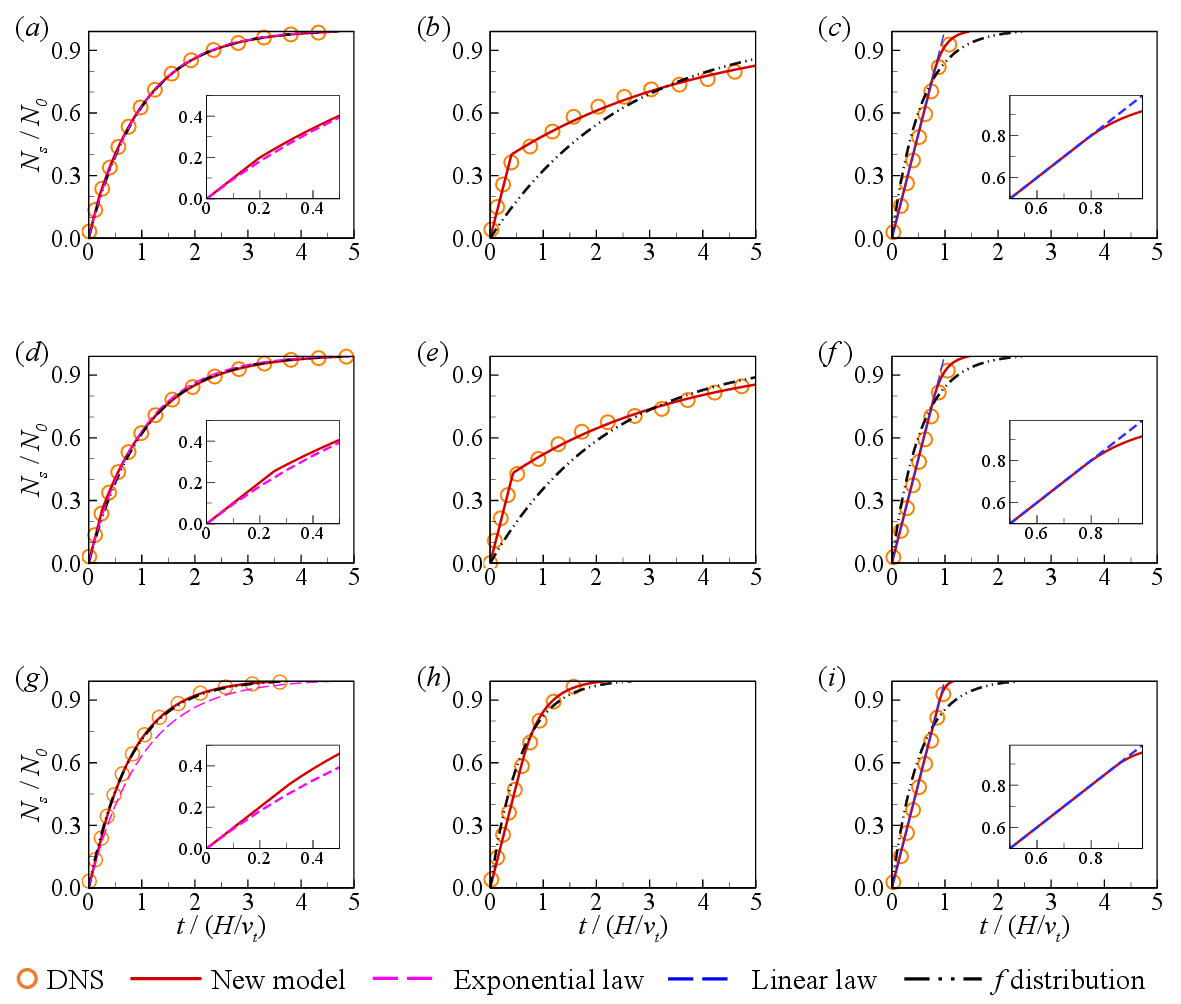}}
	\caption{Settling curves from 2-D simulations with $Ra=10^{8}$ at (\emph{a-c}) $Re_w = 0$, (\emph{d-f}) $Re_w = 2000$ and (\emph{g-i}) $Re_w = 6000$.
Particle density is fixed as ${\rho _p}= 3000$ kg/m$^3$, with diameters of (\emph{a,d,g}) ${d_p} = 5\ \mu$m, (\emph{b,e,h}) ${d_p}=20\ \mu$m and (\emph{c,f,i}) ${d_p} = 80\ \mu$m.
 The insets in (\emph{a,d,g}) magnify differences between our new model and the exponential law, and insets in (\emph{c,f,i}) magnify differences between our new model and the linear law.}
	\label{fig:settling-Model-2D}
\end{figure}

\begin{figure}
	\centerline{\includegraphics[width=0.99\textwidth]{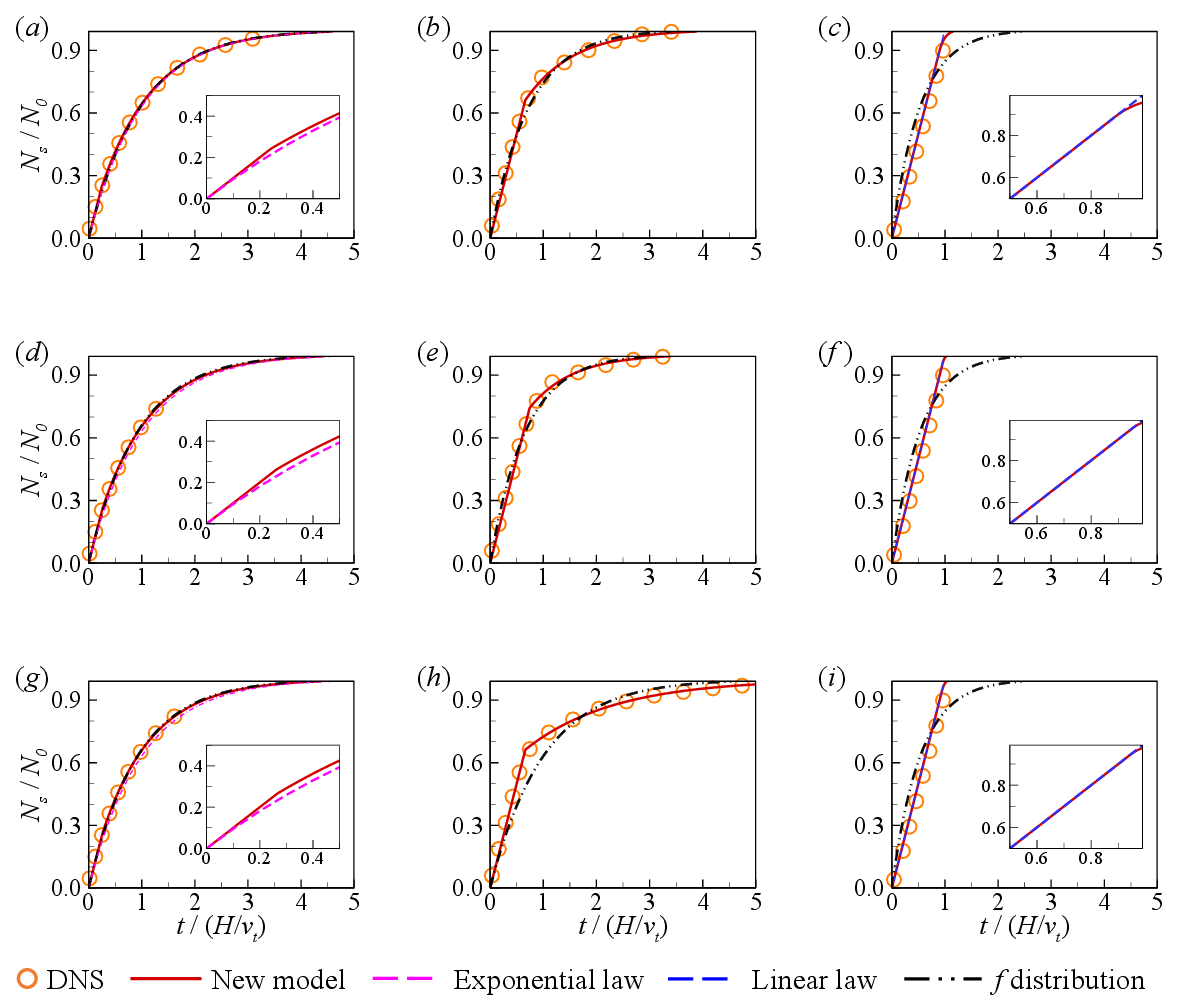}}
	\caption{ Settling curves from 3-D simulations with $Ra=10^{7}$ at (\emph{a-c}) $Re_w = 0$, (\emph{d-f}) $Re_w = 2000$ and (\emph{g-i}) $Re_w = 4000$.
Particle density is fixed as ${\rho _p}= 3000$ kg/m$^3$, with diameters of (\emph{a,d,g}) ${d_p} = 5\ \mu$m, (\emph{b,e,h}) ${d_p}=20 \ \mu$m and (\emph{c,f,i}) ${d_p} = 80 \ \mu$m.
The insets in (\emph{a,d,g}) magnify differences between our new model and the exponential law, and insets in (\emph{c,f,i}) magnify differences between our new model and the linear law.}
	\label{fig:settling-Model-3D}
\end{figure}

We quantify the deposition ratio as the ratio of settled particles $N_s$ to the initial total number $N_0$, and plot the settling curve $N_s/N_0$  from 2-D simulations as shown in figure \ref{fig:settling-Model-2D}. 
For particles with small $St$, \citet{martin1989fluid} predicted an exponential deposition rate  ${N_s}/{N_0} = 1 - \exp \left[ { - t/\left( {H/{v_t}} \right)} \right]$, and our simulation results confirmed the theoretical prediction at $Re_w \le 4000$ (see figures \ref{fig:settling-Model-2D}\emph{a,d}). 
However, departure from the expected exponential rate was observed at $Re_w = 6000$ (see figure \ref{fig:settling-Model-2D}\emph{g}), where the LSC was absent and the flow state transitioned to zonal flow. 
The main reason is that \citet{martin1989fluid} assumed convective velocities were negligible at the bottom boundary layer of the canonical convection cell; in contrast, in the strong wall-sheared convection, horizontal shear velocities persist in the boundary layer, resulting in the failure of the assumption and the deviation in the settling curve. 
For large $St$, a simple model assumes that these particles, being less influenced by the carrier flow, follow a linear deposition rate  $N_s/N_0 = t/\left( {H/{v_t}} \right)$. 
We can see from figures \ref{fig:settling-Model-2D}(\emph{c,f,i}) that our simulation results agree with the linear rate at  $t < H/{v_t}$, yet deviations are observed at the tails of the settling curve, indicating that a longer time than  $H/{v_t}$ is required for all the particles to eventually deposit on the wall. 
A possible explanation is that the carrier flow still has a subtle influence on particles with large $St$ (yet not infinitely large), and this influence becomes increasingly pronounced as time progresses. 
In figure \ref{fig:settling-Model-3D},  we further plot the settling curve $N_{s}/N_{0}$ from 3-D simulations.
Overall, the 3-D simulation results reinforce the findings from the 2-D simulations that small $St$ particles settle with an exponential deposition ratio and large $St$ particles settle with a linear deposition ratio. 

\begin{figure}
	\centerline{\includegraphics[width=0.99\textwidth]{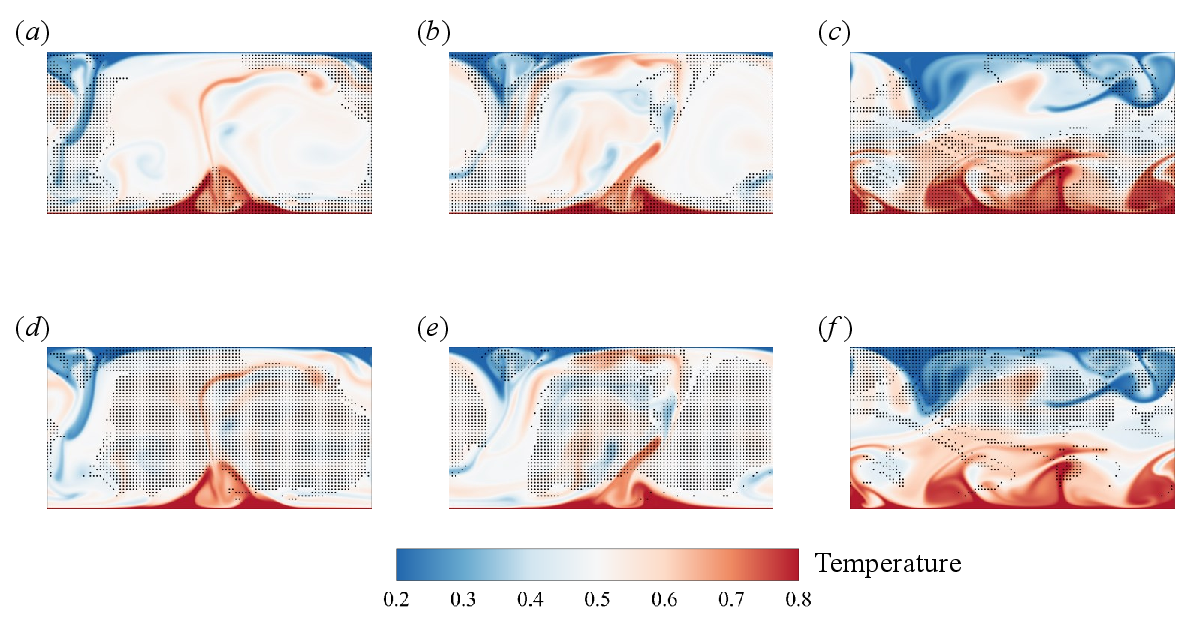}}
	\caption{An \emph{a posterior} examination of  the initial positions (black dots) for particles that have settled (\emph{a}-\emph{c}) during the linear stage and (\emph{d}-\emph{f}) during the nonlinear stage, 
alongside the corresponding instantaneous temperature fields (contours) at (\emph{a},\emph{d}) $Re_w = 0$, (\emph{b},\emph{e}) $Re_w = 2000$ and (\emph{c},\emph{f}) $Re_w = 6000$,  with $Ra=10^{8}$. 
Particle density is fixed as ${\rho _p}= 3000$ kg/m$^3$ and diameter is fixed as ${d_p}=20$ $\mu$m.
 Note this figure does not represent a 'real' instantaneous of particle distribution within the convection cell.} 
	\label{fig:LinearNonlinearPositions}
\end{figure}

\begin{figure}
	\centerline{\includegraphics[width=0.99\textwidth]{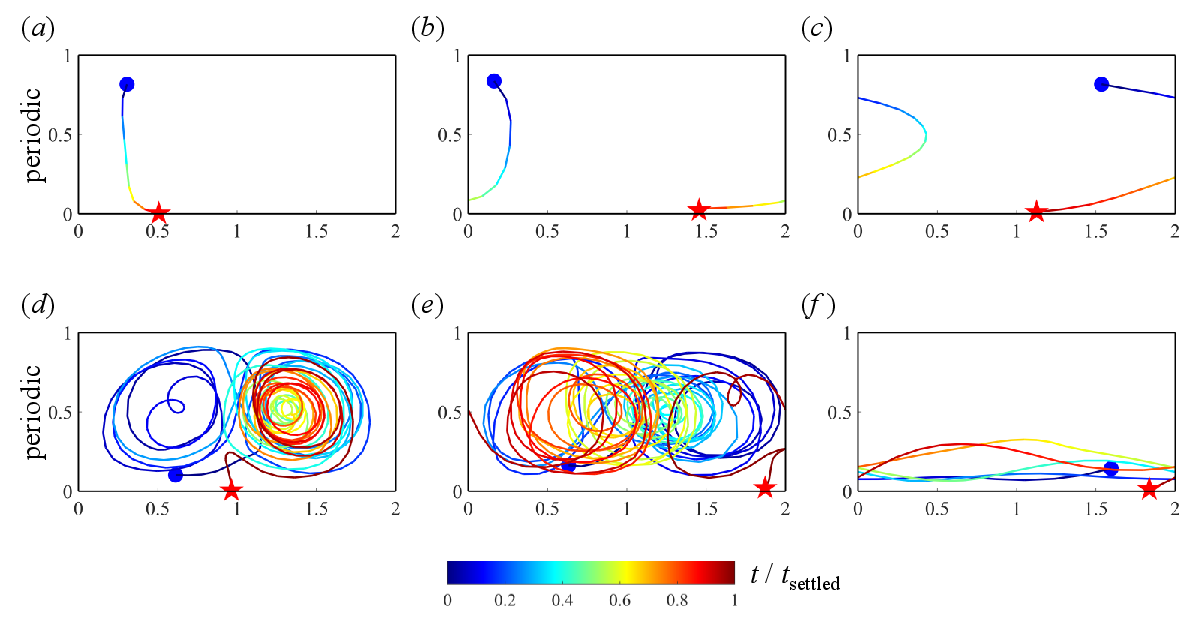}}
	\caption{Example particle trajectories for those settled  (\emph{a}-\emph{c}) during the linear stage and (\emph{d}-\emph{f})  during the nonlinear stage 
at (\emph{a},\emph{d}) $Re_w = 0$, (\emph{b},\emph{e}) $Re_w = 2000$ and (\emph{c},\emph{f}) $Re_w = 6000$,  with $Ra=10^{8}$.
Particle density is fixed as ${\rho _p}= 3000$ kg/m$^3$ and diameter is fixed as ${d_p}=20$ $\mu$m.
The trajectories are coloured according to the normalized suspension time of the particles.
 The blue circle marks the initial position and the red star marks the deposition position.}
	\label{fig:LinearNonlinearTrajectories}
\end{figure}

Despite existing simple models predicting particle deposition rates for small and large $St$, challenges arise for medium $St$ due to the complex particle-turbulence interactions. 
Examining the settling curves obtained from simulation, we develop the following model that comprises a linear settling stage succeeded by a nonlinear settling stage.
We denote by $t_{s}$ the transient time from the linear to the nonlinear stage.
At the early settling stage, particles initially released in the lower domain and in the downwelling areas  tend to settle.
To illustrate this, we conducted an \emph{a posterior} examination and marked the initial positions of particles that settle during the linear stage, as shown  in figures \ref{fig:LinearNonlinearPositions} (\emph{a}-\emph{c}). 
The particle drift equation \citep{patovcka2022residence}  gives $\mathbf{u}_{p}=\mathbf{u}_{f}+\mathbf{v}_{t}$ in the limit of  $St \ll 1$, where $\mathbf{v}_{t}$ is the particle Stokes velocity.
This equation indicates that particles subjected to updrafts settle at a speed of  ${v_t} - {v_f}$, while those in downdrafts settle at  ${v_t} + {v_f}$. 
Here,  ${v_f}$ denotes the vertical flow speed at the particle’s position. 
At this stage, the settling particles decrease height almost monotonically, without recirculating in the convection cell, as shown in figures \ref{fig:LinearNonlinearTrajectories}(\emph{a}-\emph{c}).
Assuming the same proportion of settling particles affected by the updraft  in the lower domain and downdraft  in the downwelling areas, then the numbers of $N_\text{up}$ and $N_\text{down}$ during $t \le {t_s}$ are
\begin{equation}
N_\text{up} = \frac{N_0}{2} \frac{(v_t - v_f) tLW}{HLW},
\quad
N_\text{down} = \frac{N_0}{2} \frac{(v_t + v_f) tLW}{HLW}
\end{equation}
thus, at this settling stage, the deposition ratio (i.e. the fraction of particles that have deposited on the wall relative to the initial number of particles laden in the flow system) is
\begin{equation}
\frac{N_s}{N_0} = \frac{N_\text{up} + N_\text{down}}{N_0} = \frac{t}{H / v_t}
\end{equation}
The above relation indicates that the deposition ratio linearly increases with time, which is why we name it the linear stage.

At the later nonlinear settling stage,  particles initially released in the roll centre and upwelling areas tend to settle, as illustrated in figures \ref{fig:LinearNonlinearPositions}(\emph{d}-\emph{f}).
The particles circulate within the  large-scale roll and may escape the circulation due to small-scale irregularities produced by the emergence of new plumes \citep{patovcka2020settling}. 
With the increase of wall shear, the lateral motion increases, making the particles more likely to escape the circulation, as shown in figures \ref{fig:LinearNonlinearTrajectories}(\emph{d}-\emph{f}). 
Inspired by the stochastic model proposed by \citet{denzel2023stochastic}, we denote the time duration for a particle to complete the circulation as $t_c$  and the escape probability from the circulation as  $\lambda_f$. 
The time decay rate of particle number $N$ in the convection cell is 
\begin{equation}
dN = - \lambda_f \frac{dt}{t_c} N(t) \Rightarrow N(t) = C_0 \exp \left( - \frac{\lambda_f}{t_c} t \right)
\end{equation}
thus, at the nonlinear settling stage, the deposition ratio is
\begin{equation}
\frac{N_s}{N_0} = \frac{N_0 - N}{N_0} = 1 - C_1 \exp \left( - \frac{\lambda_f}{t_c} t \right) = 1 - C_1 \exp \left( - \frac{C_2}{H / v_t} t \right)
\end{equation}
where  $C_1 = C_0/N_0$ and $C_2 = H\lambda _f/(v_t t_c)$ involves a combination of unknowns $t_c$ and $\lambda_f$.

In short, we propose a new model that describes the settling process via an early linear stage and a later nonlinear stage, written as
\begin{equation}\label{eq:Settlingmodel}
\frac{N_s}{N_0} = 
\begin{cases}
\frac{t}{H / v_t}, & t \le t_s \\
\\
1 - C_1 \exp \left( - \frac{C_2}{H / v_t} t \right), & t \ge t_s
\end{cases}
\end{equation}
Under the assumption of a smooth settling curve,  namely $N_{s}/N_{0}$ is continuously changing at $t_{s}$, we then have  ${C_1} = \left[ {1 - {t_s}/(H/{v_t})} \right]/\exp \left[ { - {C_2}{t_s}/\left( {H/{v_t}} \right)} \right]$. 
The determination of $t_s$  and $C_2$, essential for model closure, will be achieved by least square fitting of the settling curves.

Previously, in figures \ref{fig:settling-Model-2D} and \ref{fig:settling-Model-3D}, we compared results from this newly developed model with  2-D and 3-D simulation results, respectively. 
Good agreement was observed for convective flows across various $Re_w$ and particles with a wide range of $St$. 
Although our original motivation was to propose a model to predict particle deposition rates for medium $St$, the simulation results indicate that our model is applicable for small and large $St$ as well. 
In figures \ref{fig:settling-Model-2D} and \ref{fig:settling-Model-3D}, we also plot the predictions from the Pato{\v{c}}ka \emph{et al.} model (denoted as the $f$ distribution) \citep{patovcka2020settling,patovcka2022residence}. 
The key idea of the $f$ distribution is to describe particle settling as $dN/dt=-N(t)/(fH/v_{t})$, leading to a settling curve determined by $N(t)/N_{0}=\exp[-v_{t}t/(Hf)]$.
We observe a clear departure in the settling curve, mainly because their model adopts a unified exponential solution to describe the settling process, whereas DNS results indicate a linear regime in the settling curves \citep[see also in][figure 3]{patovcka2020settling}. 
As acknowledged by \citet{patovcka2020settling},  the imperfect fit of the observed settling curves is due to the fact that $f$ should be a function of time rather than remaining constant. 
This time dependency of $f$ results in a misfit between the $f$ distribution and the observed settling curves, with the largest error occurring for particles with a medium Stokes number.
Nevertheless, Pato{\v{c}}ka \emph{et al.}’s model is still useful for estimating the characteristic time of sedimentation, as discussed later in this paper.

\begin{figure}
	\centerline{\includegraphics[width=0.8\textwidth]{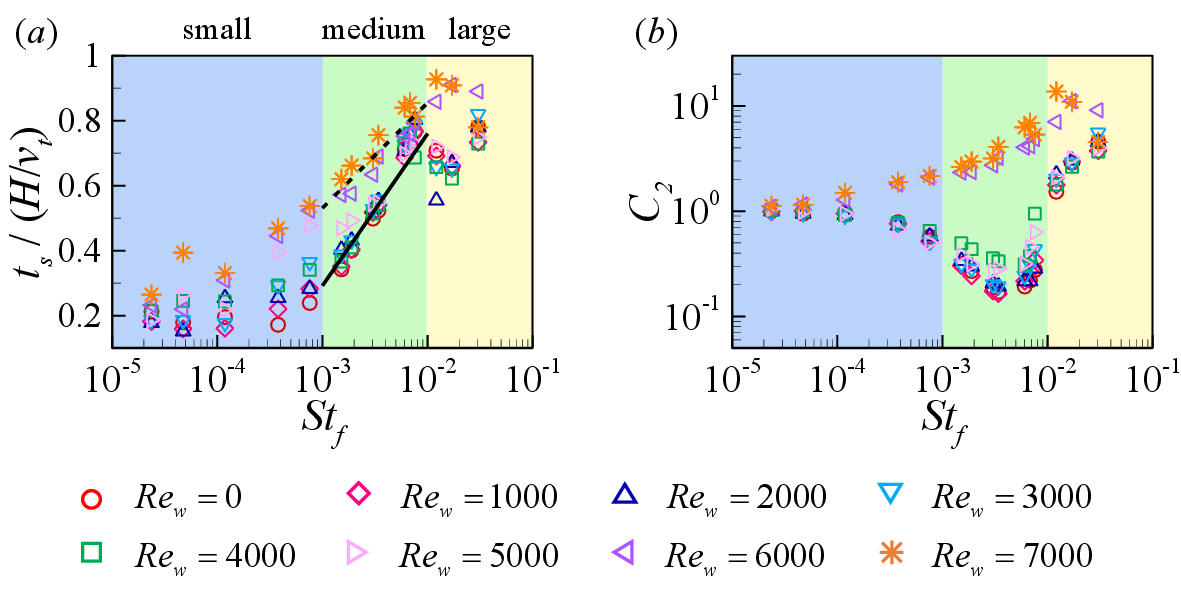}}
	\caption{Dependence of the parameters (\emph{a}) $t_s$  and (\emph{b}) $C_2$ on the Stokes number $St$ for various wall-shear Reynolds numbers $Re_w$, with $Ra=10^{8}$.
The black solid and dashed lines in (\emph{a}) represent the empirical relations (\ref{eq:ts_St}).
}
	\label{fig:Ra1e8-tsC2}
\end{figure}

For our new model to be fully closed and predictive, we still need to provide accurate modelling of the parameters $t_s$ and $C_2$. 
In figure \ref{fig:Ra1e8-tsC2}, we plot $t_s$  and $C_2$ as a function of  $St_{f}$ for various $Re_w$  in the 2-D cell with $Ra=10^{8}$.
Here, we adopt $St_f$ rather than $St_K$ as the input parameter because $St_f$ is simpler to determine from the flow parameters, making it more convenient for modelling.
In Appendix \ref{appB}  we further examine the Rayleigh number dependence of such a relationship with $Ra=10^{7}$ and $10^{9}$. 
We can see from figure \ref{fig:Ra1e8-tsC2}(\emph{a}),  for medium $St_{f}$ in the range of  $10^{-3} \le St_{f} \le 10^{-2}$, the parameter $t_s$ can be empirically estimated as
\begin{equation}\label{eq:ts_St}
\frac{t_s}{H/v_t} =
\begin{cases}
0.47 \log_{10}(St_{f}) + 1.70, & \text{if large-scale roll}  \\
\\
0.32 \log_{10}(St_{f}) + 1.50, & \text{if zonal flow}
\end{cases}
\end{equation}
This implies that $t_s$ is minorly influenced by $Re_w$ when the convective flow state remains unchanged (either in the LSC or zonal flow state). 
Practically, to empirically estimate $t_{s}$ using the above equation, we should first determine the flow states, then $t_{s}$ can be empirically expressed as a function of the Stokes number. 
This $t_{s}$ value is essentially an ensemble-averaged value for all the particles.
As for the parameter  $C_2$ (see figure \ref{fig:Ra1e8-tsC2}\emph{b}), although its original definition is just a combination of unknowns of $t_c$ and $\lambda_f$, for the convenience of mathematical calculation, in the following, we will reveal the physical meaning of this parameter  $C_2$.  

\begin{figure}
	\centerline{\includegraphics[width=0.99\textwidth]{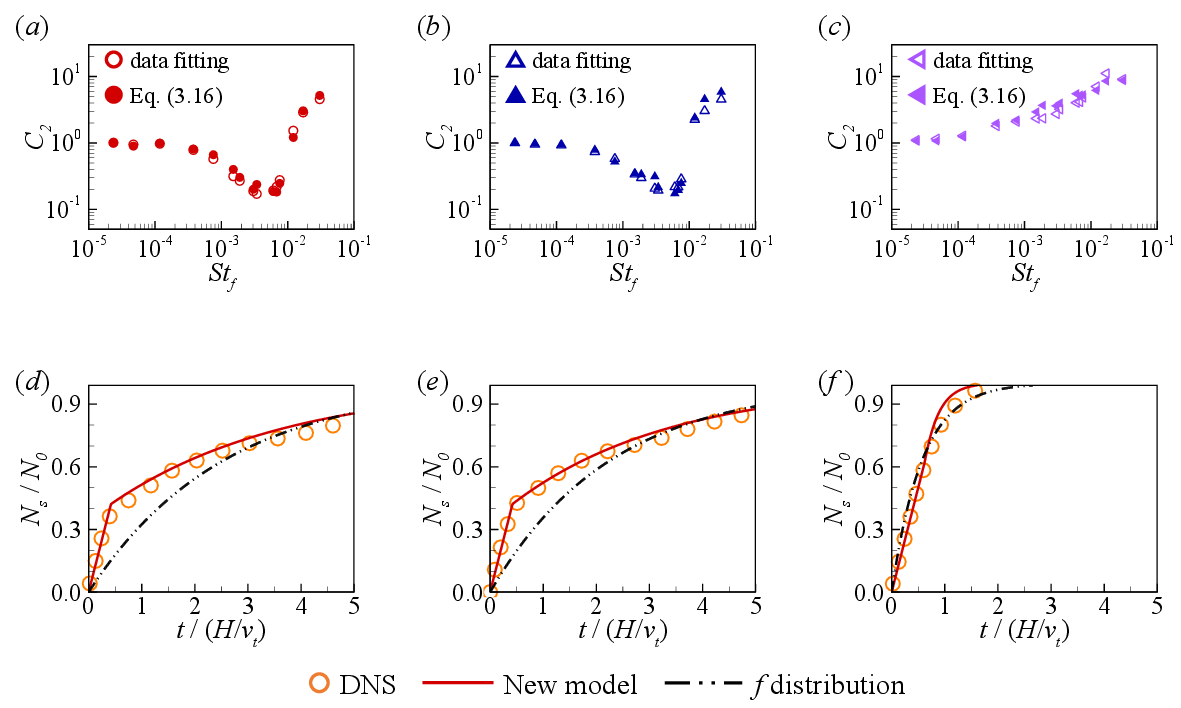}}
	\caption{(\emph{a-c}) The parameter $C_{2}$ as a function of $St$, determined from data fitting of the settling curves and  (\ref{eq:nbnd_C2}).
(\emph{d-f}) Settling curves obtained by determining unknown parameters via (\ref{eq:ts_St}) for $t_s$ and (\ref{eq:nbnd_C2}) for $C_2$, shown for (\emph{a,d}) $Re_w = 0$, (\emph{b,e}) $Re_w = 2000$ and (\emph{c,f}) $Re_w = 6000$,  with $Ra=10^{8}$. 
Particle density is fixed as $d_p$ = 3000 kg/m$^3$ and diameter is fixed as $d_p$ = 20 $\mu$m.}
	\label{fig:C2predictSettling-Ra1e8}
\end{figure}

According to the particle mass conservation equation \citep{patovcka2022residence,martin1989fluid}, the particle deposition rate $dN/dt$ is equal to the decrease of particle flux $A_0 v_t n_\text{bnd}$  at the lower boundary, namely,  $dN/dt =  - A_0  v_t  n_\text{bnd}$. 
Here, $A_0$  is the area (or length in the 2-D case) of the bottom boundary, and  $n_\text{bnd}$ is the mean particle number concentration near the lower boundary. 
Because the particle deposition rate is
\begin{equation}
\frac{dN}{dt} = 
\begin{cases}
- v_t L \frac{N_0}{LHW}, & t < t_s \\
\\
- v_t L C_2 \frac{N(t)}{LHW}, & t \ge t_s
\end{cases}
= 
\begin{cases}
- v_t L n_0, & t \le t_s \\
\\
- v_t L C_2 \left\langle n(t) \right\rangle_{V}, & t \ge t_s
\end{cases}
\end{equation}
where $\left\langle {n\left( t \right)} \right\rangle_{V}  = N(t)/(LHW)$  is the instantaneous volume-averaged concentration, then we have
\begin{equation}\label{eq:nbnd_C2}
n_\text{bnd} = 
\begin{cases}
n_0, & t \le t_s \\
\\
C_2 \left\langle n(t) \right\rangle_{V}, & t \ge t_s
\end{cases}
\end{equation}
The above analysis indicates that the parameter $C_2$ describes the ratio of mean particle number concentration near the lower boundary over the volume-averaged particle concentration during the nonlinear stage. 
From figure \ref{fig:C2predictSettling-Ra1e8}(\emph{a-c}), we observe good agreement of the $C_2$ values calculated via (\ref{eq:nbnd_C2}) with those determined by least square fitting of the settling curves. 
Then, in figure \ref{fig:C2predictSettling-Ra1e8}(\emph{d-f}), we show the settling curves by determining unknown parameters via empirical correlation (\ref{eq:ts_St}) for $t_s$ and (\ref{eq:nbnd_C2}) for $C_2$ to close the model. 
In addition to determining unknown parameters via fitting DNS data, here, we highlight that using these relations provides an alternative solution to predicting the settling curves.
Recent work by \citet{denzel2023stochastic} also relies on DNS data to determine the input statistics of their model. 
The difference is that, in \citet{denzel2023stochastic}, the input parameters for their model are determined from the DNS data as a fit to the cubic splines of the discrete p.d.f.s. for those parameters, while the input parameters for our model are determined from the DNS data as a fit to the settling curves.  
More importantly, we also demonstrate prediction of the deposition ratio using empirical correlation (\ref{eq:ts_St}) for $t_s$ and (\ref{eq:nbnd_C2}) for $C_2$ to close the model, which provides an alternative solution in addition to fitting DNS data to predict the settling curves.

\begin{figure}
	\centerline{\includegraphics[width=0.99\textwidth]{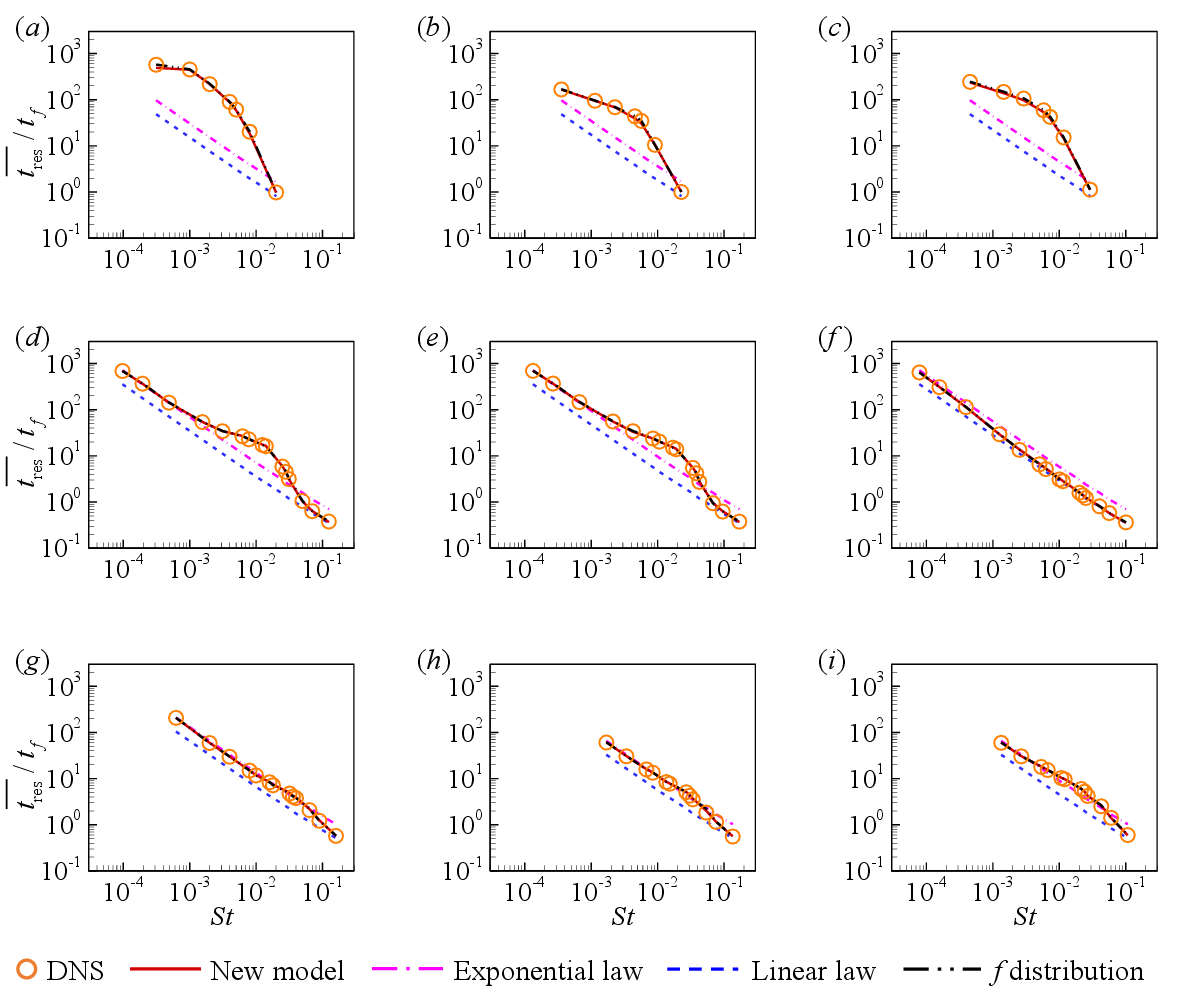}}
	\caption{ Particle mean residence time $\overline {{t_\text{res}}}$  normalized by the free-fall time  $t_f$, 
with (\emph{a}-\emph{c}) $Ra=10^{7}$, (\emph{d}-\emph{f}) $Ra=10^{8}$, (\emph{g}-\emph{i}) $Ra=10^{9}$, 
at (\emph{a}) $Re_w = 0$, (\emph{b}) $Re_w = 1000$, (\emph{c}) $Re_w = 1500$, (\emph{d}) $Re_w=0$, (\emph{e}) $Re_w = 2000$, (\emph{f}) $Re_w = 6000$, (\emph{g}) $Re_w=0$, (\emph{h}) $Re_w = 9000$ and (\emph{i}) $Re_w = 12000$.}.
	\label{fig:ResidenceTimeTf}
\end{figure}

We finally evaluate our new model in predicting the average residence time $\overline {t_\text{res}}  = \int_0^\infty  {N\left( t \right)dt} /{N_0}$, which is the duration that particles spend within the carrier fluid before deposition onto the bottom wall. 
In figure \ref{fig:ResidenceTimeTf}, we plot the average residence time $\overline {{t_\text{res}}}$, further normalized by the free-fall time  $t_f$. 
We can see that our new model is quantitatively accurate for all ranges of $St$. 
For $St\sim O(10^{-3})$, the average residence time  can exceed the free-fall time $t_f$  by up to two orders of magnitude. 
Meanwhile, we calculate the eddy turnover time as  $\tau _e = 2H/\sqrt {\left\langle {{v^2}} \right\rangle_{V,t}} $ \citep{sakievich2020temporal}, which is around  7.9 $t_{f}$ for $Ra=10^{7}$, 6.6 $t_f$ for $Ra=10^{8}$ and 5.5 $t_{f}$ for $Ra=10^{9}$ when the LSC is present.
This suggests that $\overline {t_\text{res}} $ is one order of magnitude of higher than $\tau _e$  and the particles take an average of tens of circulations within the LSC, consistent with the observation of \citet{denzel2023stochastic}. 
With the increase of $St$, the particle residence time generally decreases. 
For $St \sim O(10^{-1})$, the average residence time of a solid particle is less than the free-fall time for a fluid parcel. 
Note that, for large $St$, the particles’ residence time is significantly influenced by their initial positions and velocities; in our simulations, the particles are initially stationary and uniformly distributed. 
From these results, we can also see that the $f$ distribution is useful for estimating the mean residence time  $\overline {{t_\text{res}}}  = \int_0^\infty  {N\left( t \right)dt} /{N_0}=Hf/v_{t}$, which involves the averaged quantity of $f$ over the entire settling process. 
This averaging process smooths out the time-dependent variations, providing a good estimate of the mean residence time.

\begin{figure}
	\centerline{\includegraphics[width=0.99\textwidth]{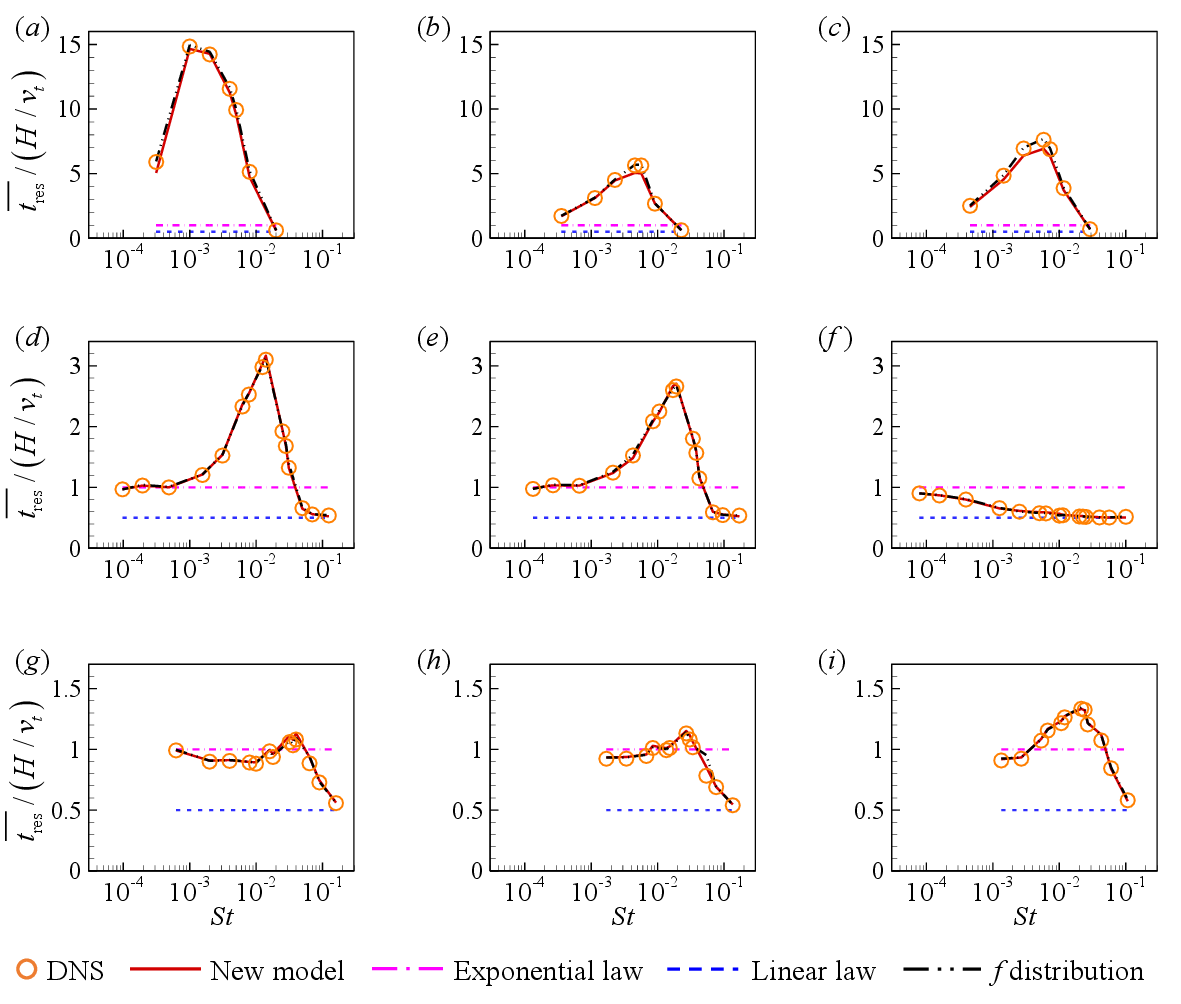}}
	\caption{ Particle mean residence time $\overline{{t_\text{res}}}$ normalized by the mean terminal time $H/{v_t}$, 
with (\emph{a}-\emph{c}) $Ra=10^{7}$, (\emph{d}-\emph{f}) $Ra=10^{8}$ and (\emph{g}-\emph{i}) $Ra=10^{9}$, 
at (\emph{a}) $Re_w = 0$, (\emph{b}) $Re_w = 1000$, (\emph{c}) $Re_w = 1500$, (\emph{d}) $Re_w=0$, (\emph{e}) $Re_w = 2000$, (\emph{f}) $Re_w = 6000$, (\emph{g}) $Re_w=0$, (\emph{h}) $Re_w = 9000$ and (\emph{i}) $Re_w = 12000$.}
	\label{fig:ResidenceTimeHVt}
\end{figure}

In figure \ref{fig:ResidenceTimeHVt}, we further plot the residence time normalized by the mean terminal time  $H/{v_t}$ as a function of $St$ for various $Re_w$. 
We compare  $\overline {{t_\text{res}}} $ with the theoretical values for particles with small $St$ and large $St$, respectively, where
\begin{equation}
\overline{t_\text{res}} = 
\begin{cases}
\int_0^\infty \exp \left( - \frac{t}{H / v_t} \right) dt = \frac{H}{v_t}, & \text{for small } St \\
\\
\int_0^{H / v_t} \frac{t}{H / v_t} dt = \frac{H}{2 v_t}, & \text{for large } St
\end{cases}
\end{equation}
For medium $St$, the particles circulate within the LSC, leading to a much longer average residence time compared with the average terminal time. 
With the increase of $Ra$, the LSC becomes more unstable \citep{zhu2018transition},  and the locking of particles inside the LSC weakens progressively \citep{patovcka2022residence}. 
For example, at a low $Ra$ of $10^7$, the average residence time can be up to ten times longer than the average terminal time, due to a fraction of particles that are trapped inside the LSC, making the average value large; 
at a high $Ra$ of $10^{9}$, the average residence time is only slightly longer than the average terminal time.
In addition, under the influence of wall shear, the LSC becomes unstable, leading to a decrease in the average residence time compared with the average terminal time. 
At $Ra=10^8$ and $Re_w = 6000$ in the 2-D simulations, the updrafts in the zonal flow are much weaker due to the absence of the uplifting force of the plumes.
Consequently, we observe faster settling of particles (see figure \ref{fig:ResidenceTimeHVt}\emph{f}). 

\section{Conclusion}
\label{sec:4 Conclusion}
In this work, we have performed DNS of wall-sheared thermal turbulence laden with point particles. 
We imposed Couette-type shear on the RB convection cell to examine the impact of interactions between vertical buoyancy and horizontal shear forces on particles’ motion. 
We observed that, with the increase of $Re_w$, the large-scale rolls expanded horizontally, evolving into zonal flow  in 2-D simulations or streamwise-oriented roll in 3-D simulations.
For particles with small and medium $St$, they either circulate within the LSC when buoyancy dominates or drift near the walls when shear dominates.
For large $St$, the turbulent flow structure has a minor influence on particles, and they settle almost similarly to those in quiescent fluid environments.

We analysed the particle distribution using Vorono\"{i} diagrams and statistical distributions of Vorono\"{i} cell areas. 
For particles with small $St$, the Vorono\"{i} cells are randomly distributed and exhibit a low degree of aggregation, which is independent of the flow state being either LSC or zonal flow. 
For medium $St$, pronounced spatial inhomogeneity and local preferential concentration are observed. 
For large $St$, clustering also occurs, but wall shear has negligible influence in contrast to cases with medium $St$.

We then derive a mathematical model for the particle concentration distribution. 
For particles with very small $St$, we obtain the averaged local particle concentration   $\left\langle {{n}\left( y \right)} \right\rangle _{t_{1/2}}/{n_0} = 1/\left( {2\ln 2} \right)$. 
For very large $St$, we have   ${\left\langle {n\left( y \right)} \right\rangle _{t_{1/2}}}/{n_0} = 2\left( {1 - y/H} \right)$ at $0.5H \le y \le H$ and ${\left\langle {n\left( y \right)} \right\rangle _{t_{1/2}}}/{n_0} = 1$  at $y \le 0.5H$. 
Compared with the DNS results, for a small but non-zero $St$, the local particle concentration generally agrees with the theoretical predictions when the LSC is present, except there is a reduced number of particles in the upper boundary layer; however, when the flow transitions to zonal flow  in 2-D simulations, significant deviation appears. 
For large but not infinite  $St$, the local particle concentration generally agrees with theoretical prediction across all $Re_w$, highlighting again that the particles' motion is nearly unaffected by the turbulent flow state.

We further plot the settling curves to quantify the deposition ratio. 
Good agreement is observed between the DNS results and previous theoretical predictions in the small and large $St$ regimes, namely, for small $St$ with an exponential deposition ratio and for large $St$ with a linear deposition ratio. 
For medium $St$, to bridge the gap, we develop a new model that describes the settling process via an early linear stage and a later nonlinear stage. 
At the early linear settling stage, particles initially released in the lower domain and in the downwelling areas settle. 
Specifically, the same proportion of particles in the updraft and downdraft regions settle with a speed of $v_t - v_f$  and  $v_t + v_f$, respectively, leading to the deposition ratio  $N_s/N_0 = t/\left( {H/{v_t}} \right)$. 
At the later nonlinear settling stage, particles circulate within the convection cell during $t_{c}$ and may escape the circulation with a probability of  ${\lambda _f}$, leading to the deposition ratio $N_s/N_0 = 1 - {C_1}\exp \left[ { - {C_2}t/\left( {H/{v_t}} \right)} \right]$. 
The unknowns in our model can be determined either by least square fitting of the settling curves, or by our empirical relations (\ref{eq:ts_St}) and (\ref{eq:nbnd_C2}). 
In addition, we found that our model for particle deposition is also accurate in predicting the average residence time across a wide range of $St$ for various $Re_w$.

\backsection[Supplementary data]{\label{SupMat}Supplementary material and movies are available at \\https://doi.org/10.1017/jfm.2024.936}


\backsection[Funding]{This work was supported by the National Natural Science Foundation of China (NSFC) through grants nos 12272311, 12125204 and 12388101, the Young Elite Scientists Sponsorship Program by CAST (2023QNRC001), and the 111 project of China (no. B17037)}

\backsection[Declaration of interests]{The authors report no conﬂict of interest.}


\backsection[Author ORCIDs]{Ao Xu, https://orcid.org/0000-0003-0648-2701;	Heng-Dong Xi, https://orcid.org/0000-0002-2999-2694}


\appendix

\section{Spatial distribution of local Kolmogorov length scale}\label{appA}

\begin{figure}
	\centerline{\includegraphics[width=0.8\textwidth]{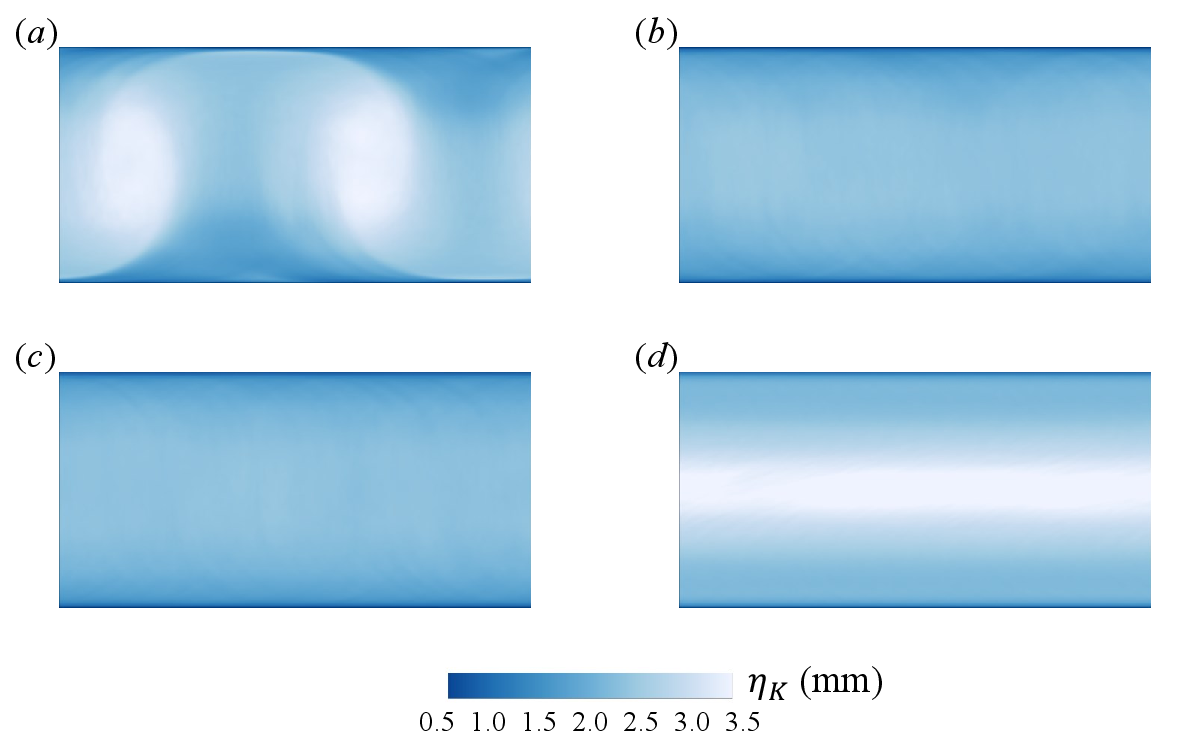}}
	\caption{ Spatial distribution of time-averaged local Kolmogorov length scale $\langle \eta_{K} \rangle_{t}$ at (\emph{a}) $Re_w = 0$, (\emph{b}) $Re_w = 2000$, (\emph{c}) $Re_w = 4000$ and (\emph{d}) $Re_w = 6000$, with $Ra = 10^{8}$ and $Pr = 0.71$ (see table \ref{tab:parameter} in the main text for the corresponding fluid properties). 
}
	\label{fig:LocalKolmogrovScale}
\end{figure}

In the main text, we presented the Kolmogorov length scale based on the volume- and time-averaged method. 
Here, we further present the time-averaged local Kolmogorov length scale distribution $\langle \eta_{K} \rangle_{t}$. 
The local Kolmogorov length scale is calculated as  $\eta_{K}(\mathbf{x},t)=[\nu^{3}/\varepsilon_{u}(\mathbf{x},t)]^{1/4}$ \citep{shishkina2010boundary}. 
Taking $Ra = 10^{8}$ as an example, we can see from figure \ref{fig:LocalKolmogrovScale}  that the minimal local Kolmogorov length scale is around 480 $\mu$m, which is six times larger than the largest particle diameter of $d_{p}=80$ $\mu$m. 
This indicates that the point-particle model can be safely used.
        
\section{Rayleigh number dependence of parameters $t_s$  and $C_2$}\label{appB}

\begin{figure}
	\centerline{\includegraphics[width=0.8\textwidth]{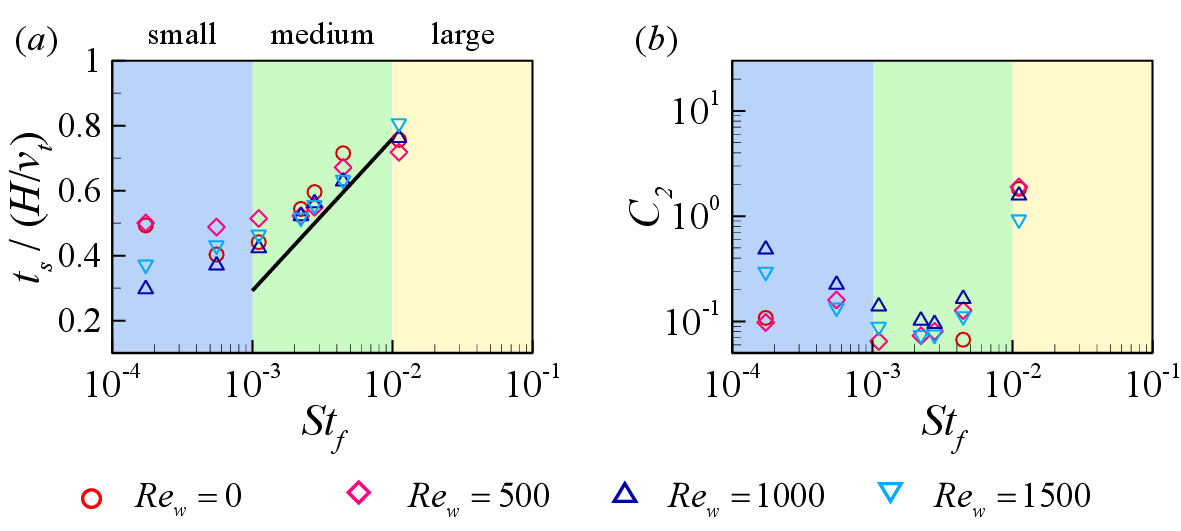}}
	\caption{ Dependence of the parameters (\emph{a}) $t_s$  and (\emph{b}) $C_2$ on the Stokes number $St$ for various wall-shear Reynolds numbers $Re_w$ with $Ra=10^{7}$.
The black solid lines in (\emph{a}) represent the empirical relations (\ref{eq:ts_St}).}
	\label{fig:Ra1e7-tsC2}
\end{figure}

\begin{figure}
	\centerline{\includegraphics[width=0.8\textwidth]{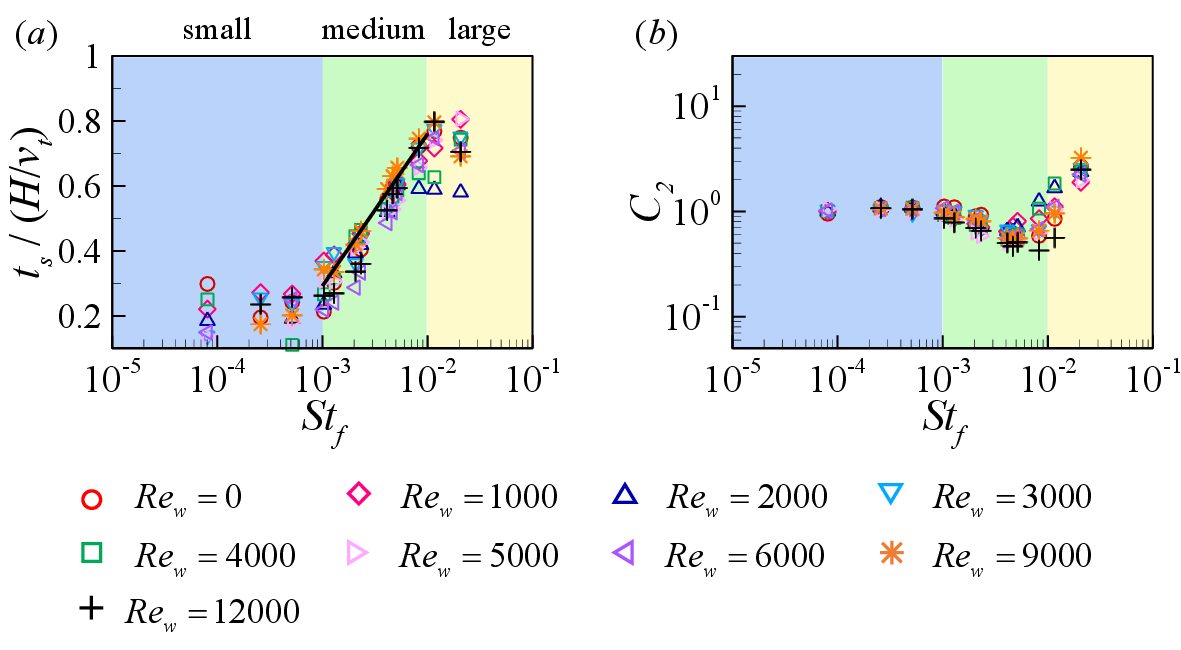}}
	\caption{ Dependence of (\emph{a}) $t_s$  and (\emph{b}) $C_2$ on $St$ for various $Re_w$ with $Ra=10^{9}$.
The black solid lines in (\emph{a}) represent the empirical relations (\ref{eq:ts_St}).}
	\label{fig:Ra1e9-tsC2}
\end{figure}

In addition to figure \ref{fig:Ra1e8-tsC2},  we further examine the Rayleigh number dependence of parameters $t_{s}$ and $C_{2}$.
In figures \ref{fig:Ra1e7-tsC2} and \ref{fig:Ra1e9-tsC2},  we present the parameters $t_{s}$ and $C_{2}$ as functions of the Stokes number $St$ for various wall-shear Reynolds numbers $Re_w$ with $Ra=10^{7}$ and $Ra=10^{9}$, respectively.
The results demonstrate that our empirical relations (\ref{eq:ts_St})  can also be applied for a wide range of $Ra$.

\bibliographystyle{jfm}

\end{document}